\documentclass{article}

\makeatletter
\@addtoreset{equation}{section}

\makeatother

\usepackage{cite}

\usepackage{graphicx}
\usepackage{amsmath}
\usepackage{amsfonts}
\usepackage{amssymb}
\usepackage[english]{babel}

\newcommand{\ef}[1]{\, #1}     

\newcommand{\be}{\begin{equation}}
\newcommand{\ee}{\end{equation}}

\newcommand{\HH}{\mathbb{H}}
\newcommand{\DD}{\mathbb{D}}
\newcommand{\R}{\mathbb{R}}
\newcommand{\de}{\partial}

\newcommand{\levy}{\mathcal{L}}

\newcommand{\osp}{\mathop{\mathfrak{osp}}\nolimits}

\newcommand{\jind}{\raisebox{-0.5pt}[0pt][0pt]{$\scriptstyle{j}$}}

\newcommand{\genus}{\mathfrak{h}}

\newcommand{\eval}[1]{\left\langle {#1} \right\rangle}

\newcommand{\Qm}{Q^{-}}
\newcommand{\Qp}{Q^{+}}
\newcommand{\Qpm}{Q^{\pm}}

\newcommand{\smfrac}[2]{\genfrac{}{}{}{1}{#1}{#2}}

\newcommand{\tr}{\mathrm{tr}}

\newcommand{\aut}{\mathrm{Aut}}

\newcommand{\vett}[1]
 {{\bf #1}}

\newcommand{\dx}[1] {\mathrm{d}{#1}}

\newcommand{\mycaption}[1]{
\parbox{11.5cm}{
\refstepcounter{figure}
\small {Figure~\arabic{figure}:} \it #1
}}

\newtheorem{proposition}{Proposition}[section]

\begin{document}

\title{\bf Spanning Forests on Random Planar Lattices}

\author{Sergio Caracciolo, Andrea Sportiello,\\
\small{Dip.~di Fisica dell'Universit\`a degli Studi di Milano  and INFN,}\\
\small{via Celoria 16, I-20133 Milano, Italy}\\
\small{\tt 
Sergio.Ca{}racciolo@{}mi.inf{}n.it,
Andrea.Spo{}rtiello@{}mi.in{}fn.it
}}

\date{March 24, 2009}

\maketitle

\begin{center}
{\it
    Dedicated to \'Edouard Br\'ezin and Giorgio Parisi, \\ 
    pioneers also in this subject, \\
    on the occasion of their special birthday.
}
\end{center}

\smallskip

\begin{abstract}
\noindent
The generating function for spanning forests on a
lattice is related to the $q$-state Potts model in a certain $q
\to 0$ limit, and extends the analogous notion for spanning trees, or
dense self-avoiding branched polymers.  Recent works have found a combinatorial
perturbative equivalence also with the (quadratic action) $O(n)$ model
in the limit $n \to -1$, the expansion parameter $t$ counting the
number of components in the forest.
\newline
We give a random-matrix formulation of this model on the ensemble of
degree-$k$ random planar lattices. For $k=3$, a correspondence is
found with the Kostov solution of the loop-gas problem, which arise as
a reformulation of the (logarithmic action) $O(n)$ model, at $n=-2$.
\newline
Then, we show how to perform an expansion around the $t=0$ theory.  In
the thermodynamic limit, at any order in $t$ we have a finite sum of
finite-dimensional Cauchy integrals. The leading contribution comes from a
peculiar class of terms, for which a resummation can be performed
exactly.
\end{abstract}

\vspace{2mm}





\section{Introduction}

The $O(n)$-invariant $\sigma$-model for $n=0, -1, -2$ defined on a
generic graph $G$ has a very interesting combinatorial interpretation.

Already in early 70's, it has been observed that the  $n$-vector model
in the limit in which $n\to -2$ is equivalent to a free fermionic
theory~\cite{Balian-Toulouse} . But the quadratic (Gaussian) term is
the Laplacian on the graph $G$ and the partition function is its
determinant, which according to Kirchhoff matrix-tree theorem,
provides, once the zero-mode has been removed, the weight of spanning
trees of the graph $G$~\cite{dupdav}.

In 1980 Parisi and Sourlas~\cite{Parisi-Sourlas} showed the
equivalence of the $n$-vector model in a limit in which $n\to 0$, which
was already known to describe the critical behaviour of
polymers~\cite{deGennes}, with a supersymmetric $\osp(2|2)$ model in
which the loops in Feynman graphs, which vanish in the $n\to 0$ limit,
gives zero contribution, because of the cancellation between bosons
and fermions (independently also McKane noticed that fermions can be
used to cancel the contribution of bosonic loops~\cite{McKane}). Also
this construction is independent from the choice of the graph $G$, as
only exploits the symmetry properties in the target space. The choice
for the $O(n)$-invariant model as a $\sigma$-model provides exactly
the partition function of self-avoiding walks on the graph (see for
example~\cite{ACF}).

More recently, we have shown~\cite{noi} that the generating function
of spanning forests in a graph $G$ can be represented as a Grassmann
integral of the exponential of a local fermionic weight involving a
Gaussian term together with a special nearest-neighbour four-fermion
interaction (see also~\cite{noi_hyper}). Furthermore, the fermionic
model possesses a hidden $\osp(1|2)$ supersymmetry non-linearly
realised.  In~\cite{noi} we also discussed briefly how this fermionic
model can be mapped, at least in perturbation theory, onto an
$\osp(1|2)$-invariant $\sigma$-model with spins taking values in the
unit supersphere in $\mathbb{R}^{1|2}$, or an $O(n)$-invariant
$\sigma$-model with spins taking values in the unit sphere in $\R^n$
(also known as \emph{$n$-vector model}), analytically continued to
$n=-1$. The parameter $t$ which appears in the generating function of
the forests to count the number of trees in a forest is related to the
coupling constant in the $\sigma$-model (with an important inversion
of sign). It is remarkable that the same generating function can be
obtained by a suitable limit $q\to 0$ of the $q$-state Potts model
defined on the same graph $G$ (see for example~\cite{Alan}).

Very detailed information on these models can be obtained by
considering {\em regular} graphs, and in particular in two dimensions,
where methods of Conformal Field Theory and Integrable Systems apply.

The critical exponents of the $n$-vector model, at least on the range
$n\in [-2,2]$, can be computed {\em exactly}, thanks to a mapping onto
the solid-on-solid model~\cite{nienhuis} of a suitable choice of the
weights, which, in the high-temperature expansion forbids loop
crossing. We shall call this variant of the $n$-vector model the
\emph{Nienhuis model}, or \emph{Loop-gas model} (as it is a `hard-core
lattice gas' in which the elementary objects are self-avoiding loops).
Also the critical behaviour of the Potts model can be analyzed
exactly, thanks to the mapping onto an ice-type model, which has been
constructed both algebrically~\cite{TL} and
combinatorially~\cite{Baxter}. This is once more mapped to a
solid-on-solid model~\cite{Beijeren}. And the critical limit of the
solid-on-solid model is recovered by using the Coulomb-gas
picture~\cite{DG}. More precisely, the critical behaviour of this
$n$-vector model is given by a conformal field theory (CFT) with
central charge
\begin{equation}
c(n) \,=\, 1-\frac{6}{m(m+1)} \label{cn}
\end{equation}
where the parameter $m$ is related to $n$ by the relation
\begin{equation}
n \,=\, 2 \cos \frac{\pi}{m}
\end{equation}
and to the Coulomb-gas coupling constant by
\begin{equation}
g_0 \, =\, 1 + \frac{1}{m}\, .
\end{equation}
Please, remark that $c(-1)=-3/5$ and $c(-2)=-2$. This means that the
Nienhuis model does not describe at $n=-1$ the universality class of
spanning forests. 
Indeed, a direct perturbative analysis of the $O(n)$ $\sigma$-model on
a square lattice~\cite{sergio} or on a triangular
lattice~\cite{claudia} at $n=-1$ shows that the model is
asymptotically free for $t=0^+$. As a result, the ultra-violet fixed
point is the free theory which describes trees, therefore, aside
logarithmic violations, the central charge is $c=-2$. 

Beyond the study of a model on a fixed periodic planar graph, a lot of
progress has been achieved by considering an ensemble of planar
graphs~\cite{thooft,Brezin}.  Such a study has both an interest \emph{per se}
of combinatorial nature, and a relevance in connection to the original
case of regular graphs.  Indeed, after the work of Knizhnik, Polyakov
and Zamolodchikov~\cite{polya30, kpz}, it is now understood that, for systems
showing conformal invariance at criticality, statistical averages in
the two ensembles
have related critical behaviours, so that informations in one context
can be inferred from the other, e.g.~concerning the critical
exponents, the conformal families of operators, and their dimensions.
\cite{polya30, kpz, David88a, DistlerKawai88, duplungo}
In many cases, cross-checks of these predictions have been performed
\cite{DK2, DK, duplungo}.

A deep understanding of KPZ relation is a hard and active field
\cite{David88a, dupKPZ, DavidNew}.
However, some heuristic reasons can be given at least for the
existence of a relation of this kind.  
In two dimensions, at criticality, scale invariance, combined with (discretized) Euclidean symmetries, is promoted to the symmetry described by the full Virasoro Algebra.  
Analogously, statistical mechanics
models on random planar graphs, when reaching simultaneously the
large-volume limit and the critical point for the ``matter fields''
(\emph{double scaling limit}), show the scale invariance pertinent to
criticality, combined with the (discretized) invariance under local
diffeomorphisms (as bare random planar graphs describe a
discretization of two-dimensional quantum gravity). 
But a conformal theory in presence of two-dimensional quantum gravity enjoys a symmetry 
corresponding to a
$\mathrm{SL}(2, \mathbb{R})$ current algebra~\cite{polya30}, which is
larger than that described by the Virasoro algebra.


This richer structure is in a way at the root of the fact that many
results exist for the apparently harder counting problem on random
planar graphs (which involves a double average), and still lacks in
the Euclidean case (and indeed provide a hint to critical aspect of
these quantities, through KPZ).  In words more appropriate to the
discrete setting, in many combinatorial approaches (among which the
present paper), the interplay between degrees of freedom of the
lattice and of matter fields plays a crucial role in simplifying the
expressions. On the other side, many Euclidean concepts involving a
natural notion of distance are harder to define cleanly in the
random-graph setting (and, if defined through geodesic distance, hard
to compute).


On the other side, the generating function of statistical
configurations over random graphs (as well as many other ``global''
physical observables, such as susceptibilities) can be written as the
Feynman expansion of a proper action of a zero-dimensional field
theory: the replacement of real or complex bosonic fields with $N
\times N$ symmetric- or hermitean-matrix fields allows to count graphs
of genus $\genus$ with a weight proportional to $N^{-2\genus}$, and a
large-$N$ limit, achieved via steepest descent or continuous
approximation of matrix spectra, gives the restriction to planar
graphs. Such a strategy, started with the seminal works of 
\cite{thooft, Brezin},
had a strong development in the subsequent thirty years, and
now deserves the name of \emph{Random Matrix} technique
(see~\cite{difranc} for a recent pedagogical introduction).
 
Indeed, after the Ising~\cite{Kazakov} and Potts~\cite{kazakov_potts}
models, also the Nienhuis model has been solved~\cite{kostov, Gaudin} on
random planar graphs, and deeply studied (see for example 
\cite{eynardzinnjustin, Kristjansen, koststau}).
In~\cite{SA}, as we are interested in the cases $n=-1,
-2$ of the Nienhuis model, a combinatorial reformulation of these
problems has been introduced to achieve the random matrix solution
with no need of an analytical continuation.

A detailed account of these different research areas when they overlap
on geometrical critical phenomena is given by Duplantier and
Kostov~\cite{DK}.

But the model of spanning forests in an ensemble of random graphs
escapes the realm of exact results. Only the limit of spanning trees,
that is $t=0$, that we have seen corresponds to the Nienhuis model at
$n=-2$, has been studied, and for regular graphs with coordination
number 3. For spanning forests it is necessary to dispose of
informations at finite values of the coupling constant, at least in
perturbation theory.

A full discussion on the contents of this paper is postponed to the
end of section 2, after that some other definitions are introduced.

\section{The model}
\label{sec.model}

Since here on, and according to common use in Random
Matrix, we call a ``graph'' $G$ (or, more precisely, a ``fatgraph''),
what is indeed an orientable 2-dimensional cell complex, 
i.e.~$G$ is determined not only by the sets $V(G)$ and $E(G)$, 
respectively for vertices and edges, but also by a consistent choice
of the set $F(G)$ for the \emph{faces} (or, equivalently, for each
vertex, the outgoing edges have a given cyclic ordering). For this
reason, the \emph{genus} $\genus$ of $G$ is univocally and easily
determined, for example via Euler formula ($V$, $E$, $F$ and $K$
denote respectively the number of vertices, edges, faces and connected
components in the cell complex)
\be
\label{eq.Euler}
2 \genus = 2K+E-V-F
\ef.
\ee

Consider the ensemble of all connected graphs with vertices of degree
$k$, with a measure depending from the genus (and allowing to take a
``planar'' limit).  Most commonly studied cases are $k=3$ of $k=4$: we
will study the generic case, but with special attention to $k=3$, both
because it is the ensemble studied in Kostov solution, and therefore
this allows for a direct comparison of results, and because the
generating function $A_1(\omega)$ for cubic trees is the one 
with the simplest explicit formula.


For a connected graph $G$ with $V$ vertices and genus $\genus$ we
consider the customary (unnormalized) measure for Random Matrix theory
\be
\label{eq.measRG}
\mu_{g,N}(G)=\frac{1}{|\aut(G)|} g^{V} N^{-2 \genus}
\ef,
\ee
and the generating function (for connected graphs) is obtained 
by the Random Matrix technique with a single matrix field:
\be
Z_0=\sum_G \mu_{g,N}(G) =
\frac{1}{N^2} \ln \int_{N \times N}\!\!\!\!\!\!\!\! \dx M
\,e^{N \tr \left( -\smfrac{1}{2} M^2 + \smfrac{g}{k} M^k \right)}
\ef.
\ee
Here the integral is over a set of $N^2$ real variables, arranged
in the Hermitian matrix $M$, and must be intended as the formal
Feynman expansion in the parameter $g$~\cite{difranc}. Indeed, a sketch of the
technique is the following: the Feynman diagrammatics
leads to the generating function of all connected graphs, and the traces due
to the ``matrix of fields'', jointly with the Wick rule
$\eval{M_{ij} M_{\ell k}}=\delta_{j \ell} \delta_{k i}$, 
lead to a combinatorics on face-indexing that, via
Euler formula, allows to count the genus of the graph with the desired factor
$N^{-2 \genus(G)}$.

This is a \emph{one-matrix theory}, i.e.~we have only one matrix of
fields. We recall here some results.  First perform the change of
variables $M \to U \Lambda U^{-1}$, with $U$ unitary and $\Lambda$
diagonal, the Jacobian corresponding to the square Vandermonde
determinant $\Delta^2(\vec{\lambda})= \prod_{i \neq j} |\lambda_i -
\lambda_j|$.  Angular degrees of freedom do not appear in the action,
and are trivially integrated, similarly the ordering of the
eigenvalues is irrelevant, and is trivially summed over, and the
partition function reads
\be
\label{eq.Zpuregrav}
Z_0 \propto
\frac{1}{N^2} \ln \int \!\dx{{}^N\!\vec{\lambda}}
\, \Delta^2(\vec{\lambda})
\, \exp \Big[ N
\sum_i\left( -\smfrac{1}{2} \lambda_i^2 + \smfrac{g}{k} \lambda_i^k
\right)
\Big]
\ef.
\ee
At this point, many tools allow to extract the 
asymptotic behaviour in $V$ in the planar limit (saddle point,
orthogonal polynomials, loop equations\ldots). The result for the
leading behaviour near to the radius of convergence $g_c$ of the
series is, for any genus $\genus$,
\begin{align}
Z_0(g, \genus) 
&\sim 
c(\genus) 
\sum_{V}
\left( \frac{g}{g_c} \right)^V 
V^{-3+\gamma+ \genus \gamma'}
N^{-2 \genus}
\ef,
&
\gamma
&=
-\frac{1}{2}
\ef;
&
\gamma'
&=
\frac{5}{2}
\ef.
\end{align}
The quantity in (\ref{eq.Zpuregrav}) is the \emph{pure gravity}
partition function, which should be adopted as the appropriate
normalization factor when comparing with the analogous expression in
the case of a matrix theory describing an ensemble of combinatorial
structures on the graph (interpreted as \emph{matter fields} coupled
to the gravity). In our case, we should deal with Fortuin-Kasteleyn
random clusters for the analytic continuation in $q$ of the $q$-colour
Potts Model, in a limit $q \to 0$ corresponding to a restriction
to forests, with a factor $t$ per tree in the forest.

So, let $\mathcal{F}(G)$ be the set of spanning
forests over the graph $G$. Say for short that $F \prec G$ if $F \in
\mathcal{F}(G)$, and call $K(F)$ the number of its connected
components. Given two graphs $G$, $H$ with $H \subseteq G$, define $G
\diagup H$ the \emph{contraction} of $G$ by $H$, i.e.~the graph in
which vertices $i$ and $j$ are identified if an edge $(ij)$ is in
$E(H)$ (this is the concept used, for example, in deletion/contraction
operations for Tutte-Grothendieck invariants).  In particular, if $G$
has $V$ vertices and $E$ edges, and $F \prec G$ has $E'$ edges, the
graph $G \diagup F$ has $V-E'$ vertices and $E-E'$ edges. Furthermore,
as any tree is homotopic to a point, $G$ and $G \diagup F$ have the
same topology ($\genus(G) = \genus(G \diagup F)$), however they may
have different automorphism groups.

At fixed graph $G$,
we define the spanning-forest generating function as
\be
Z(t;G)=\sum_{F \prec G} 
\mu_{t,G}(F)
\ef,
\ee
through the (unnormalized) measure
\be
\mu_{t,G}(F) = t^{K(F)} 
\frac{|\aut(G)|}{|\aut(G \diagup F)|}
\ef.
\ee
The symmetry factor, depending on the pair $(G,F)$, has been
introduced for later convenience. Here we remark that, if we worked
with \hbox{(leg-)labeled} graphs, and exponential generating function,
i.e.
\be
Z=\sum_{G~\textrm{labeled}} \frac{1}{(2 |E(G)|)!} \mu(G)
\ef,
\ee
the rephrasing of the measure above would involve no factors
whatsoever.  Furthermore, for large graphs (which are the dominant
class in the thermodynamic limit), these symmetry factors are almost
surely~1.

A remarkable exception to this last argument occurs in our exact
resummation of section \ref{sec.pertHO}, where the contributions of
certain families of ``large-volume'' graphs are resummed into the
evaluation of certain integrals over graphs with size of order 1, for
which the inclusion of the proper symmetry factor is important.

We are interested in the macrocanonical average in the random lattice
ensemble, mainly in the limits $N \to \infty$ (\emph{planar} limit),
and $g \to g_c(t)$, the radius of convergence of the resulting series
(\emph{thermodynamic} limit). A key fact is that interchanging the two
sum operations (over graphs $G$, and over forests $F$ on $G$) the
expression largely simplifies. Indeed, easy manipulations give
\be
\begin{split}
Z(t,g,N)
& = \sum_G \mu_{g,N}(G) Z(t;G) 
= \sum_G \sum_{F \prec G} 
\mu_{g,N}(G) \mu_{t,G}(F)
\\
& = \sum_G g^{|V(G)|} N^{-2 \genus(G)} \sum_{F \prec G} 
\frac{t^{K(F)}}{|\aut(G \diagup F)|}
\\ 
& = \sum_F t^{K(F)} g^{|V(F)|} \sum_{G \succ F} 
\frac{N^{-2 \genus(G \diagup F)}}{|\aut(G \diagup F)|}
\ef,
\end{split}
\label{eq.Zfor}
\ee
and in particular, for the planar case $N \to \infty$, we have
$N^{-2 \genus(G)} \to \delta_{\genus(G), 0}$, and
\be
Z(t,g)
= \sum_F t^{K(F)} g^{|V(F)|} 
\sum_{\substack{G \succ F \\ \textrm{planar}}} 
\frac{1}{|\aut(G \diagup F)|}
\ef.
\label{eq.Zforplan}
\ee
Estimating this function as well as possible, in the neighbourhood of
$t \sim 0$ and $g \to g_c(t)$, that is, for graphs in the thermodynamic
limit, and near to the critical point of spanning trees,
is the main goal of this paper.

In section \ref{sec.precomb} we give some preliminary results of
combinatorial nature.
In section \ref{sec.RMfor} we describe a standard random-matrix
approach to the problem at generic $t$, $g$ and $N$. In sections
\ref{sec.nien} and \ref{sec.ON}, it is shown a correspondence with the
$O(n)$ loop-gas model.

In sections \ref{sec.perttrees}--\ref{sec.pertHO}, the function
(\ref{eq.Zforplan}) is shown to admit
a diagrammatic perturbative expansion 
in the parameter $t$, i.e.~the coefficients $Z_n(g)$ of the series
\be
\label{eq.Zforpert}
Z(t,g) = \sum_{n\geq 1} t^n Z_{n}(g)
\ef 
\ee
can be calculated with a combinatorial technique, each term being a
finite sum of finite-dimensional integrals.

The way in which these diagrams are effectively evaluated is described
in section \ref{sec.feyneval}, while sections
\ref{sec.nonana} and \ref{sec.pansy} 
describe how to extract the
leading behaviour from this sum, in the double limit $t \to 0$ and $g
\nearrow g_c$, for $t/ \ln (g_c/g)$ below a critical threshold.

While sections \ref{sec.RMfor}, \ref{sec.nien} and \ref{sec.ON}
involve concepts of Random Matrix Theory, all other sections
are purely of combinatorial nature, and use elementary
self-contained methods (except for the isolated appearence of the
non-trivial L\'evy generalized Central Limit Theorem, in
section~\ref{sec.pertF2}, which furthermore is not mandatory in the
logic of the paper, as the results are rederived in
section~\ref{sec.feyneval}).


\section{Preliminary combinatorial results}
\label{sec.precomb}

Here we make a short review of results for certain counting problems
which will arise in our analysis. This section does not make use of
any tool from Random Matrix theory, and is fully elementary and
self-contained.
We call $\HH$ the upper half plane,
and $\DD$ the unit disk.

\begin{figure}[t]
\begin{center}
\setlength{\unitlength}{25pt}
\begin{picture}(9.5,6.2)
\put(0,0){\includegraphics[scale=1., bb=5 250 260 400, clip=true]{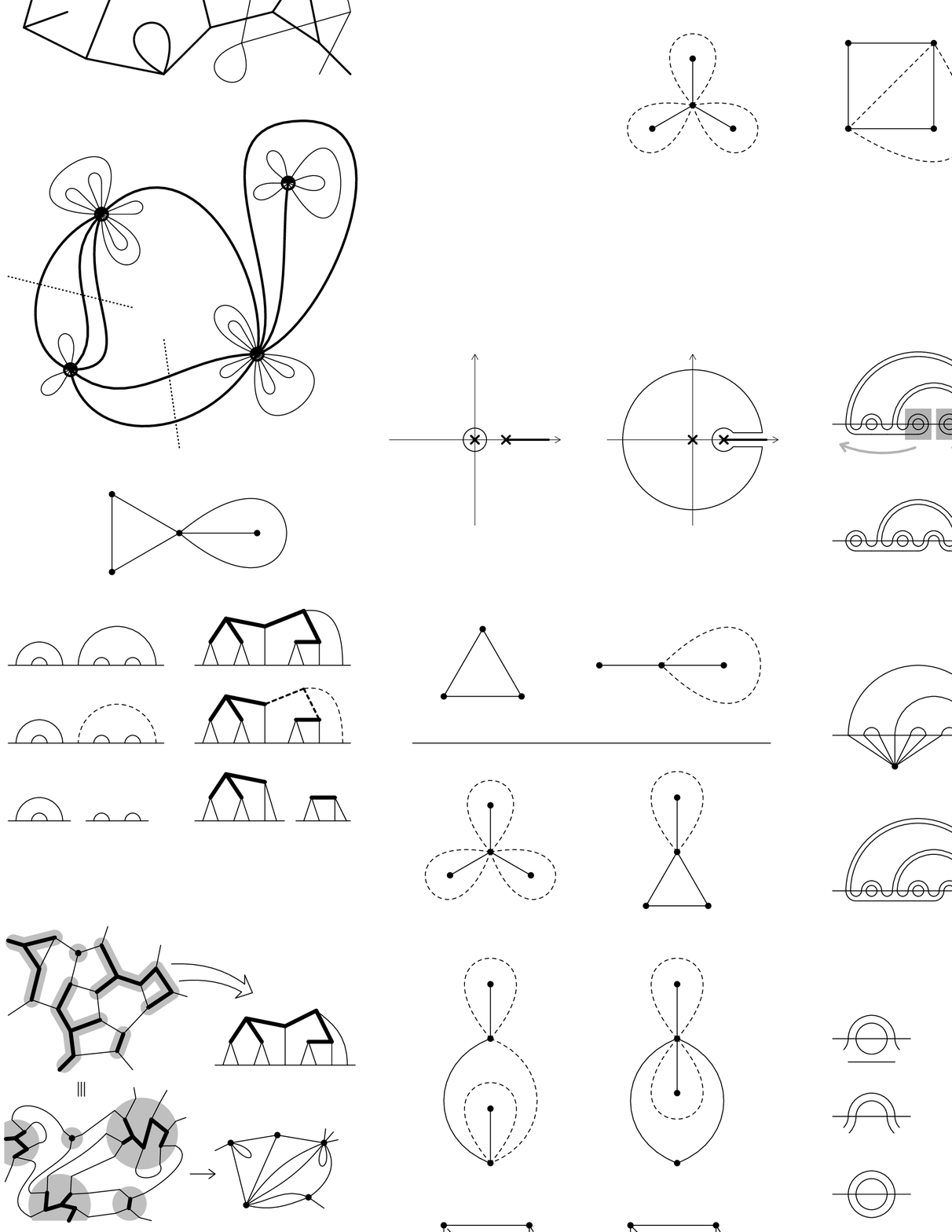}}
\end{picture}
\mycaption{\label{fig.comb}On the left, a link pattern on $\HH$ and the removal
  procedure which leads to the relation (\ref{eq.reccatal}). On the
  right, a cubic tree on $\HH$, and the removal procedure which leads
  to the relation (\ref{eq.a1}).}
\end{center}
\end{figure}

\vspace{1mm}
\noindent
{\bf Problem 1.}~{\it Counting the configurations of $n$ non-intersecting
arcs in $\HH$, with endpoints in $\de \HH$
(cfr.~fig.~\ref{fig.comb}, top left).}
\vspace{1mm}

\noindent
Such a configuration is called a \emph{link pattern} (in $\HH$). Their
number is $C_n$, the $n$-th Catalan number. We adopt the convention
$C_0=1$ for the empty configuration.  Call $C(q{})=\sum_n C_n q{}^n$
the generating function.  A sketch of proof is as follows.  For $n
\neq 0$, one can remove the right-most arc. Then, the inner and outer
part of the original diagram correspond to smaller independent
configurations, of sizes $n'$ and $n-n'-1$. This leads to a
convolutional relation for the coefficients $C_n$, and a polynomial
one for the generating function
\be
\label{eq.reccatal}
C(q{})=1 + q{}\, C(q{})^2
\ef,
\ee
the solution matching the regularity condition $C_{-1}=0$ being
\be
\label{eq.catalgf}
C(q{})=\frac{1-\sqrt{1-4 q{}}}{2 q{}}
\ef.
\ee
The closed expression for the Catalan numbers is
\be
\label{eq.catal}
C_n=\frac{1}{n+1} \binom{2n}{n}
\ef.
\ee

We say that a tree has \emph{degree $k$} if all of its vertices have
degree either $k$ or $1$, the latter being called \emph{leaves}
(indeed, any tree with at least one edge must have at least two
leaves). The word \emph{cubic} is used here as a synonimous of
\emph{degree 3}.
We recall that, for a tree to have $v_d$ vertices of degree $d$,
the set of $v_{d}$'s must satisfy the constraint 
\be
\sum_d (2-d) \, v_d = 2
\ef.
\ee
Then we will consider the problem

\vspace{1mm}
\noindent
{\bf Problem 2.}~{\it Counting the configurations of non-intersecting
  cubic trees, with $n$ vertices of degree 3 in $\HH$ and $n+2$ leaves
  in $\de \HH$ (cfr.~fig.~\ref{fig.comb}, top right).}
\vspace{1mm}

\noindent
Call $A_{1,n}$ the number of such trees, and $A_1(q{})$ the generating
function $A_1(q{}) = \sum_{n \geq 0} A_{1,n} q{}^n$.  We have
$A_{1,0}=1$ for the tree with a single edge and two vertices of degree
1.  For $n>0$, the right-most leaf must be connected to a vertex of
degree 3. Then, we can modify the drawing in a unique way so that each
of the two other terminations can be seen as the right-most leaf
vertex of the remaining component. Thus we have the formula
\be
\label{eq.a1}
A_1(q{})=1 + q{} A_1(q{})^2
\ef,
\ee
i.e.~we deal again with Catalan numbers,
\be
\label{eq.catalgfQQQQ}
A_1(q{}) =C(q{})=\frac{1-\sqrt{1-4 q{}}}{2 q{}}
\ef,
\ee
and $A_{1,n}=C_n$. 

We will now consider a generalization to arbitrary degree $k$:

\vspace{1mm}
\noindent
{\bf Problem 3.}~{\it Counting the configurations of non-intersecting
  trees of degree $k=h+2$, with $n$
  vertices of degree $k$ in $\HH$ and $hn+2$ leaves in $\de \HH$.}
\vspace{1mm}

\noindent
The reasoning follows as above, with the only difference that now,
under removal of the right-most leaf and the corresponding internal
vertex, we have $h+1$ remaining components, thus we have
\be
\label{eq.defAh}
A_h(q{})=1 + q{} \big( A_h (q{}) \big)^{h+1}
\ef,
\ee
which gives a generalization of Catalan numbers~\cite{sloane}
\be
\label{eq.adef}
A_{h,n}=\frac{1}{hn+1} \binom{(h+1)n}{n}
\ef.
\ee
Performing a Stirling expansion of (\ref{eq.adef}) for large $n$,
we get
\be
\label{eq.adefSt}
A_{h,n} \simeq
\left( \frac{(h+1)^{h+1}}{h^h} \right)^n
n^{-\frac{3}{2}}
\sqrt{\frac{h+1}{2 \pi h^3}}
\ef.
\ee
Conversely, equation (\ref{eq.defAh}) for the generating function for
general $h$ cannot be written in a simple form. 
A futher exception (besides $h=1$) is the case
$h=2$, for which we have
\be
A_2\Big( \frac{4 x^2}{27} \Big)
= \frac{3}{x} \sin \left( \frac{1}{3} \arcsin x
\right)
\ef.
\ee
In Appendix
\ref{app.HG} it is shown that the solution to (\ref{eq.defAh}) is a generalized
hypergeometric function ${}_{h+1}F_h$, and some properties are
studied.  Here, we discuss a minimal set of properties which are
strictly necessary in the forthcoming sections.

Defining for future convenience the combination
\be
\label{eq.gcrit_pre}
g_c(h)= 2^{-h} \frac{h^h}{(h+1)^{h+1}}
\ef,
\ee
we see that the radius of convergence for $A_h(q{})$ is
\be
\label{eq.convrad}
|q{}| < 2^h g_c(h)
\ef.
\ee
The series has a finite value also for this value, indeed
\be
A_h \big( 2^h g_c(h) \big) = \frac{h+1}{h}
\ef,
\ee
and, as the series has positive summands, \emph{a fortiori}
$\left| A_h \big( 2^h g_c(h) e^{i \theta} \big) \right| \leq 
\frac{h+1}{h}$.

Equation (\ref{eq.defAh})
has a parametric solution in algebraic form
\be
\label{eq.solparam}
\left\{
\begin{array}{l}
q{}^{1/h} = z = x (1-x^h) \ef; \\
A_h= (1-x^h)^{-1} \ef;
\end{array}
\right.
\ee
which is easily checked by direct substitution, and by matching the
initial condition $A_{h,0}=1$.
By a simple scaling we have, for $z=x(1-g x^h)$,
\begin{align}
\label{eq.solparamMoregen}
A_h(g z^h) 
&= (1- g x^h)^{-1} 
\ef;
&
z \big( A_h(g z^h) - 1 \big)
&= g x^{h+1}
\ef.
\end{align}
The problem of counting non-intersecting trees in $\HH$ is obviously
related to the one of counting non-intersecting trees in $\DD$ (with
leaves in $\de \DD$).  The solutions essentially do coincide (if one
point on $\de \DD$ is marked), except for the fact that, for the case
on the disk, it is natural to add a symmetry factor for the cyclic
permutations of the leaves. Thus we introduce the modified quantities
\begin{align}
\label{eq.a'def}
A'_{h,n}
&=\frac{A_{h,n}}{hn+2}
\ef;
&
A'_h(q{})=\sum_{n\geq 1} q{}^n A'_{h,n}
\ef.
\end{align}
Remark that for $A'$, differently than for $A$, we start summation
from $n=1$.
Again the case $k=3$, i.e.~$h=1$, gives a simple explicit formula
\be
\label{eq.a'1}
A'_1(q{})
=
\frac{-1+6 q{} - 6 q{}^2 + (1 - 4 q{})^{\frac{3}{2}} }
{12 q{}^2}
\ef.
\ee
More generally, the definition (\ref{eq.a'def}) implies the relation
\be
\label{eq.aa'}
z \left( A_h(g z^{h}) -1 \right)
= \frac{\mathrm{d}}{\mathrm{d}z }
\left( 
z^2 A'_h(g z^h) \right)
\ef,
\ee
which, 
using
$\frac{\mathrm{d}}{\mathrm{d}z} f(z(x)) 
= ( \mathrm{d}z / \mathrm{d}x )^{-1}
\frac{\mathrm{d}}{\mathrm{d}x} f(z(x))$,
gives
\be
\label{eq.30jhgtris}
z^2
A'_h(g z^h)
=
\frac{x^2}{2} 
\left( \frac{2}{h+2} g x^{h} - (g x^{h})^2 \right)
\ef.
\ee

\section{Random-matrix partition function for spanning forests}
\label{sec.RMfor}

Conside the set of pairs $(G,F)$ where $G$ is a connected graph with
all vertices of degree $k=h+2$, and $F\in \mathcal{F}(G)$.  With our
choice for the measure, the partition function $Z(t,g,N)$ is given by
(\ref{eq.Zfor}).
Call $\{ T_{\alpha} \}_{\alpha = 1, \ldots, K(F)}$ the components of
the forest.
Consider the edges $E(F) \subseteq E(G)$ as \emph{marked}, and the
remaining edges as \emph{unmarked}.
Conversely, consider \emph{all} vertices as marked. Now the
quantity $K(F)$ coincides with the number of connected marked
components (possibly consisting of isolated vertices).
We say that an unmarked edge is an \emph{arc} if it connects
points on the same component, and a \emph{bridge} if it connects
points on distinct components.
\label{page.bridgesarcs}

For each edge $(ij) \in E(G)$, call \emph{leg-decoration} the
introduction (in series) of two intermediate vertices, $i_j$ and
$j_i$:
\[
\setlength{\unitlength}{20pt}
\begin{picture}(8,0.7)
\put(0,0){\line(1,0){3}}
\put(0,0){\circle*{.25}}
\put(3,0){\circle*{.25}}
\put(5,0){\line(1,0){3}}
\put(5,0){\circle*{.25}}
\put(6,0){\circle*{.25}}
\put(7,0){\circle*{.25}}
\put(8,0){\circle*{.25}}
\put(-0.2,0.4){$i$}
\put(2.8,0.4){$j$}
\put(4.8,0.4){$i$}
\put(5.8,0.4){$i_j$}
\put(6.8,0.4){$j_i$}
\put(7.8,0.4){$j$}
\put(3.6,0.1){$\longrightarrow$}
\end{picture}
\]
Call \emph{leg} a decorated edge of the form $(i\, i_j)$. This
decoration provides a language for the mechanism of Wick contractions
underlying the Random Matrix technique: the vertices coming
from the expansion of the action determine the set of legs (with their
cyclic ordering), while the choice of Wick contractions determines
how the legs are connected.

Leg-decorate all unmarked edges.  For each component $T$ of $F$, we
define its \emph{border} as the set of unmarked legs incident on
$T$. As $G$ is actually a 2-dimensional cell complex, and any tree is
homotopic to a point, the legs on the border of a tree $T$ have an
induced cyclic ordering.

Consider a component $T$ with $n$ vertices, together with its $hn+2$
border legs, as a new tree $T'$ with $n$ vertices all of degree $k$,
and $hn+2$ leaves.  We know from problem 3 that the number of such
configurations is exactly $A_{h,n}$,
and, if divided by the factor $hn+2$ given by the cyclic symmetry, 
is $A'_{h,n}$.

\begin{figure}[t]
\begin{center}
\setlength{\unitlength}{25pt}
\begin{picture}(9.5,8.2)
\put(0,0){\includegraphics[scale=1., bb=0 -5 260 200, clip=true]{figs.eps}}
\end{picture}
\mycaption{\label{fig.reduct}On top, a portion of a typical configuration 
of spanning forests on a random 3-graph, and a planar cubic tree
corresponding to the right-most connected component. Below, on the
left, a
manipulation of the drawing which highlights the shape of the
``effective vertices''; on the right, the diagram of the one-matrix partition
function which contains the contribution of the original
configuration, obtained shrinking the effective vertices to single
points, by keeping memory only of the external-leg sequence.}
\end{center}
\end{figure}

Imagine to contract these trees $\{ T'_{\alpha} \}$ to single
vertices, without altering the cyclic ordering of the external
legs. Only edges of the form $(i_j \, j_i)$ survive, and the
combination of all possible $n$-vertex trees, with $n \geq 1$, leads
to an ``effective'' coupling $g_{n}$ for the resulting diagram, which
coincides with $G \diagup F$.
Recalling that we have a factor $t$ per component, and that we
introduced the proper symmetry factor in the definition of $A'_{h,n}$,
we should write
\be
\frac{g_{n}}{hn+2} = 
t g^n A'_{h,n}
\ef.
\ee
The combinatorics of the resulting (fat)\-graphs $G \diagup F$ is then
suitable for re\-summation
through Random Matrix technique.
Introduce a Hermitian matrix of fields $M$, and
consider
the action
\be
\begin{split}
\mathcal{S}(M)
&=\tr 
\bigg( -\frac{M^2}{2} 
+ \sum_{n \geq 1} \frac{g_{n}}{hn+2} M^{hn+2} \bigg)
\\
& =
-\frac{1}{2} \tr \left( M^2 (1 - 2 t A'_{h}(g M^{h}) \right)
\ef,
\end{split}
\ee
this leads to our desired generating function:
\be
Z(t,g,N)=\frac{1}{N^2} \ln
\int_{N \times N}\!\!\!\!\!\!\!\! \dx M
\,e^{N \mathcal{S}(M)}
\ef.
\ee
A graphical explanation of the whole procedure is presented in
figure~\ref{fig.reduct}. 

So, we still deal with a one-matrix theory, as in the case of
pure-gravity, although, unfortunately the ``potential'' is not
polynomial in matrix fields (and this causes problems, for example, in
the solution methoud through the resolvent function).  Again we can
reduce to eigenvalues, and obtain
\begin{gather}
Z(t,g,N)=\frac{1}{N^2} \ln
\int_{N}\!\! \dx{\vec{\lambda}}\, \Delta^2(\vec{\lambda})
\, e^{-N \sum_i V(\lambda_i)}
\ef;
\\
V(\lambda)= \frac{\lambda^2}{2} (1 - 2 t A'_h(g \lambda^{h}))
\ef.
\end{gather}
where integration over values of $\lambda$ is intendend inside the
analiticity region for $V(\lambda)$, that is, using the result of
(\ref{eq.convrad}),
\be
| \lambda_i | < 
2 \left( \frac{g_c(h)}{g} \right)^{\frac{1}{h}}
\ef.
\ee
We can reduce the potential to a polynomial, via the proper change
of variable, inspired by relation (\ref{eq.30jhgtris})
\be
\lambda_i(x_i)=x_i (1- g x_i^{h})
\ef;
\ee
such that the interesting quantities change into
\begin{gather}
\dx{\lambda_i} = \dx{x_i} (1- g (h+1) x_i^{h})
\ef;
\\
\begin{split}
&| \lambda_i - \lambda_j |
=
| (x_i-x_j)-g (x_i^{h+1}-x_j^{h+1}) |
\\
&\quad=
| x_i-x_j | \cdot
| 1 - g( x_i^{h} + x_i^{h-1}x_j + \cdots + x_j^{h}) |
\ef;
\end{split}
\\
V(x)=\frac{x^2}{2} \left(
1- 2 g \left(1+\smfrac{t}{h+2}\right) x^{h} 
+ g^2 \left(1+t \right) x^{2h} \right)
\ef.
\end{gather}
Remark how the corrections to the measure and to the Vandermonde factor
combine, to give the factor
\be
\widehat{\Delta}_g (\vec{x})
=
\prod_{i, j} ( 1 -g (x_i^{h} + x_i^{h-1}x_j + \cdots + x_j^{h} ))
\ef;
\ee
such that now the partition function reads
\be
\label{eq.Zfinal}
Z=\frac{1}{N^2} \ln
\int_{N}\!\! \dx{\vec{x}}\, \Delta^2(\vec{x})
\widehat{\Delta}_g (\vec{x})
\, \exp \Big(-N \sum_i V(x_i) \Big)
\ef.
\ee

\section{A remark on the $O(n)$ loop-gas model}
\label{sec.nien}

Given the spherical $O(n)$ probability measure
\be
\dx{\mu(\vec{\sigma})} =
\frac{2}{ \Omega_n}\, \delta(\sigma^2-1) \, \dx{{}^n \sigma} 
\ef,
\ee
with $\Omega_n = \frac{2 \, \pi^{\frac{n}{2}} }{ \Gamma\left(\frac{n}{2}\right)}$, 
the lowest moments are given by
\begin{align}
\label{eq.momon}
\eval{\sigma^a}
&=
\eval{\sigma^a \sigma^b \sigma^c}= 0
\ef;
&
\eval{\sigma^a \sigma^b}= \frac{1}{n} \delta_{ab}
\ef.
\end{align}
Consider a graph $G$ in which each vertex has degree at most 3:
the action
\be
S_G = 
\sum_{(i,j) \in E(G)}
\ln (1 + n \beta \; \vec{\sigma}_i \cdot \vec{\sigma}_j )
\ef 
\ee
is such that the polynomial expansion of the integrand in the
partition function involves only the moments in (\ref{eq.momon}). More
specifically, the terms of the expansion of
\be
\prod_{(i,j) \in E(G)}
(1 + n \beta \; \vec{\sigma}_i \cdot \vec{\sigma}_j )
\ee
are in correspondence with the configurations 
$\vec{a} \in (\{0 \} \cup \{ 1, \ldots, n \})^{E(G)}$, where an unmarked edge
$(ij)$ (that is, $a_{ij}=0$) corresponds to a choice of the summand 1
in the expansion of factor $(ij)$, and an edge marked with colour $a$ 
(that is, $a_{ij}=a$) corresponds to the choice of summand
$n \beta \, \sigma^a_i \sigma^a_j$.

Integration over the spherical measure at each vertex leaves only with
configurations of marked self-avoiding loops, weighted with a factor
$\beta$ per marked edge. Summing over loop colourings also produces
the ``topological'' factor $n$ per loop \cite{nienhuis}.

The same result would have been obtained for any spherical measure
such that equations (\ref{eq.momon}) hold, i.e., up to a rescaling,
for any spherical measure. We remark however that, if we choose to
integrate the variables with the very special function such that not
only equations (\ref{eq.momon}) hold, but also all higher momenta vanish, the
combinatorial $O(n)$-model--loop-gas correspondence extends to
\emph{generic} graphs and actions, up to the possible appearance of
dimers. Indeed, assume the action has a series expansion
\be
\label{eq.S5676}
S=\sum_{\eval{ij}} \sum_{k \geq 1}
\beta_k n^k ( \vec{\sigma}_i \cdot \vec{\sigma}_j )^k
\ef.
\ee
As momenta higher than $\eval{(\sigma^a)^2}$ vanish, all the
coefficients $\beta_k$ with $k \geq 3$ are irrelevant. We can
equivalently parametrize $\beta_{1,2}$ as
\begin{align}
\beta_1
&=\beta
\ef;
&
\beta_2
&=
-\smfrac{1}{2} \beta^2 + \smfrac{1}{n} \gamma
\ef;
\end{align}
and the integrand of the partition function is
\be
e^{S(\sigma)}=
\prod_{\eval{ij}} 
\bigg(1 + n \beta \sum_a \sigma^a_i \sigma^a_j
+ n \gamma \sum_a (\sigma^a_i \sigma^a_j)^2 \bigg)
+ 
R( \sigma )
\ef,
\ee
where the remainder term $R( \sigma )$ is a polynomial in the fields,
in which each monomial has at least either a factor $\sigma_i^3$ or a
factor $\sigma^a_i \sigma^b_i$ with $a \neq b$, and thus vanishes
after integration, with our choice for the invariant measure.
So we can consider a combinatorics of unmarked (summand
$1$), marked with colour $a$ (summand $n \beta \, \sigma^a_i \sigma^a_j$)
and doubly-marked with colour $a$ 
(summand $n \gamma (\sigma^a_i \sigma^a_j)^2$) edges. Variable
integration produces $1$ if all adjacent edges are unmarked, $1/n$ if
two adjacent edges are marked, and with the same colour, or one edge is
doubly-marked, and 0 otherwise. So we are left with configurations of
coloured self-avoiding loops and dimers, edges in the loops being
weighted with a factor $\beta$, and dimers with a factor $\gamma/n$,
Summing over possible colourings reproduces the ``topological'' factor
$n$ per loop, and rescales the weight of marked (but uncoloured) dimers to
$\gamma$. The case $\gamma=0$ in the action (\ref{eq.S5676}), on a
cubic lattice, corresponds to the loop-gas problem studied in the
literature, in \cite{nienhuis} and subsequent works.

\section{Connection between the spanning-forest and the $O(n)$
  loop-gas model}
\label{sec.ON}

The problem of counting configurations of self-avoiding closed loops,
with a weight $\beta^{L} n^{\ell}$ given to a configuration with
$\ell$ loops of total length $L$, is expected to be a combinatorial
variant of the $O(n)$ model, in particular in the case $k=3$, where
the combinatorial derivation is more transparent, and for this reason
has been widely studied first by Kostov, and afterwords by many
others, with a large number of interesting 
results~\cite{kostovMPL89, Gaudin, eynardzinnjustin, Kristjansen}.

Here we briefly sketck the derivation, in order to highlight the
similarities with our partition function, in equation
(\ref{eq.Zfinal}). It turns out that a stronger analogy emerges in the
generalization of the loop-gas problem considered in section
\ref{sec.nien}, so we will study the  
problem of counting configurations of
self-avoiding closed loops and dimers, on random graphs of
coordination $k=h+2$,
with a weight $\beta^{L} n^{\ell} \gamma^d$ given to a
configuration with $\ell$ loops, of total length $L$, and $d$
dimers.

One can study the problem with a random-matrix technique, introducing
$n+2$ Hermitian
matrix fields, a matrix $M$ for unmarked edges, a matrix $A$ for
the dimers, and $n$
auxiliary matrices $\{ E_{\alpha} \}_{\alpha=1, \ldots, n}$, 
which mark the edges of the loops in one of $n$ colors, 
in order to reproduce the ``topological'' factor
$n^{\ell}$.
Thus we have the random-matrix partition function
\begin{gather}
Z_{\textrm{l-d gas}}=\frac{1}{N^2} \ln
\int \dx{(M, A, \{ E_{\alpha} \})}
e^{N \mathcal{S}(M, A, \{ E_{\alpha} \})}
\ef;
\\
\begin{split}
\mathcal{S}
&= \tr \Big[
-\frac{1}{2} \Big( M^2 + A^2 + \sum_{\alpha} E_{\alpha}^2 \Big)
+\frac{g'}{k} M^k 
\\
&
+ \gamma^{\frac{1}{2}} g' A M^{k-1}
+\sum_{\alpha, h'}
\frac{\beta g'}{2} M^{h'} E_{\alpha} M^{h-h'} E_{\alpha}
\Big]
\ef.
\end{split}
\end{gather}
If we perform the Gaussian integration of matrix $A$, we have
\be
\mathcal{S}
= \tr \Big(
-\frac{1}{2} (M^2 + \sum_{\alpha} E_{\alpha}^2)
+\frac{g'}{k} M^k + \gamma {g'}^2 M^{2k-2}
+\sum_{\alpha, h'}
\frac{\beta g'}{2} M^{h'} E_{\alpha} M^{h-h'} E_{\alpha}
\Big)
\ef.
\ee
Consider the change of variables
\begin{align}
M 
& \to U \Lambda U^{-1}
\ef;
&
E_{\alpha} \to U E_{\alpha} U^{-1}
\ef;
\end{align}
where the Jacobian is just the one-matrix 
Vandermonde determinant, as the
transformation is unitary on the auxiliary matrices $E_{\alpha}$.
Now, for any pair $i \leq j$, each term 
$(E_{\alpha})_{ij}$ appears in the action with a quadratic
contribution proportional to
\be
(E_{\alpha})_{ij} (E_{\alpha})_{ji}
\Big( 1- \beta g' (\lambda_i^{h} + \lambda_i^{h-1} \lambda_j + 
\cdots + \lambda_j^{h}) \Big)
\ee
and thus
the Gaussian integration of $E_{\alpha}$ degrees of freedom gives
a factor $\widehat{\Delta}^{-n/2}_{\beta g'}(\vec{\lambda})$
\begin{gather}
\label{eq.Zfinal675476}
Z_{\textrm{l-d gas}}=\frac{1}{N^2} \ln
\int_{N}\!\! \dx{\vec{x}}\, \Delta^2(\vec{x})
\widehat{\Delta}^{-n/2}_{\beta g'}(\vec{x})
\, \exp \Big(-\sum_i V(x_i) \Big)
\ef;
\\
V(x) = \frac{1}{2} x^2 - \frac{g'}{k} x^{k} - \gamma {g'}^2 x^{2k-2}
\ef.
\end{gather}
If analitic continuation in $n$ can be performed, at least
for negative integers values of 
$n$~\footnote{Cfr.~\cite{SA} for a discussion on this point.},
in the case $n=-2$ we recover the partition function
(\ref{eq.Zfinal}) up to constants, with the parameter correspondence
\be
\left\{
\begin{array}{l}
g'/g = 1/\beta = h+2+t \ef; \\
\rule{0pt}{4mm}\gamma 
 = -\frac{1}{2} \, (g'/g)^{-2} \, (1+t) \, g^{\frac{2}{h}} \ef;
\end{array}
\right.
\ee
the only last delicate point being the fact that in 
section \ref{sec.RMfor} we
required all $x_i$ to be in the region of analicity of the function
$A'_h$, while in this case we require that all the factors
$(1-\beta g' (h+1)x_i^h)$ are strictly positive, in order to make
positive-definite the Gaussian integration of the auxiliary degrees of
freedom (clearly, $1-a(h+1)x_i^h>0$ for all $i$ implies
$1-a\sum_{h'} x_i^{h'} x_j^{h-h'} >0$ for all pairs $(i,j)$). One
should check that these requirements coincide. A one-line argument is
that the quantity $\ln \widehat{\Delta}(\vec{x})$ acts as a sort of 
``repulsive potential'' from the border of the allowed domain, thus,
as it coincides in the two cases, also the borders of the two domains
must coincide.

For $h=1$, the case of Nienhuis loop-gas, with no dimers, is obtained
for $\gamma=0$, that is at $t=-1$. More generally, at arbitrary
degree, a loop-gas model with vertex-disjoint loops and no dimers is
still obtained for $t=-1$. We remark that this value of the coupling
is special in the theory: the generating function of spanning forests
is `probabilistic' only for real positive values of $t$, by which we
mean that each configuration takes a real positive weight, which can
be interpreted as an (unnormalized) Gibbs measure. For $t$ real
negative, this picture does not hold anymore, in the combinatorial
formulation in terms of forests. It does hold, however, for an
alternate description, in terms of spanning trees only, weighted with
some non-local factors related to ``activities'', this being the
original Tutte description of the generating function. In particular,
for the generic Random Cluster model we have the definition of
\emph{internal} and \emph{external} activities, while specialization
to spanning forests gives unitary weight to externally-active edges
(so that it is not necessary to count them), and gives a factor
$(1+t)$ per internally-active edge. So, in the interval $t \in
[-1,+\infty)$, the description of the generating function in terms of
trees and activities is probabilistic.  Tutte's notion of activity
requires a choice of a linear order on the edge set (though the
generating function of the activities is, in fact, independent of this
order).

Various other definitions of notions of activities exist, leading to
the same Tutte generating function, in a non-obvious way. In
particular, an astonishingly different notion has been recently
introduced by Olivier Bernardi \cite{bernardiTutte}. His notion only
requires a cyclic ordering of the edges around each vertex (and,
again, the resulting generating function is, in fact, independent of
this choice). Not only it is interesting that, for a generic graph,
this characterization strongly restricts the range of arbitrariness in
the accessory ordering structure, but also, at our purposes here, that
it is specially natural for graphs which are already embedded on the
Riemann sphere, as is for the ensemble of ``random planar graphs''
2-dimensional cell complexes arising from Random Matrix theory. In
this case, the embedding naturally defines such a cyclic ordering, and
no arbitrariness whatsoever is required in the Bernardi construction
of Tutte polynomial (except for the choice of a single starting
directed edge). So, this is a natural candidate for the construction
of a probabilistic combinatorial expansion of our generating function,
and the investigation of the model at $t \searrow -1$, when internal
activities are forced to vanish. This combinatorial approach could
shed a light on the natural conjecture that the model is critical at
$t=-1$ for any graph degree, and in the universality class of Nienhuis
loop-gas model.

\section{Perturbative expansion: spanning trees}
\label{sec.perttrees}

In the previous sections we sketched how to deal with the
model at generic $t$ with standard random-matrix techniques, even at
finite $N$ (and, for example, have access to higher-genus generating
functions via loop equations).

Beside this, a purely combinatorial approach based on the results of
section \ref{sec.precomb} is enough to determine the planar partition
function perturbatively in $t$ at any given order, with a relatively
small effort (also higher-genus quantities could be calculated with
this technique, but this is not discussed here). This approach is
interesting, not only by itself, and for the emerging combinatorics,
but also for a comparison with the analogue perturbative calculations
on flat 2-dimensional lattices.

In this section we perform the first-order calculation, concerning
spanning trees, while in sections from \ref{sec.pertF2} to 
\ref{sec.pansy}
we will describe the higher-order technique.

So, we deal with the problem of counting the pairs $(G,T)$, with $G$ a
connected planar graph with $V$ vertices, all of degree $k=h+2$, and
$T$ a spanning tree on $G$, denoted by the symbol $T \prec G$. We have
\be
Z_1(g)= \sum_T g^{V(T)} \sum_{G \succ T} \frac{1}{|\aut(G \diagup T)|}
\ef.
\ee
The graph $G \diagup T$ contains only one vertex, of coordination
$hV+2$. The edges form a link pattern connecting these terminations.

The combinatorics of the coefficients, together with the constraint
that $hV$ is even (the ensemble of coordination-$k$ graphs
with $V$ vertices is empty if both $V$ and $k$ are odd, as there are
no pairings of an odd number of legs), produces
\be
Z_1(g)= 
\left\{
\begin{array}{ll}
\displaystyle{\sum_V g^V A'_{h,V} C_{(hV+2)/2}} & \textrm{$h$ even;}
\\
\rule{0pt}{14pt}%
\displaystyle{\sum_{\textrm{$V$ even}}
       g^V A'_{h,V} C_{(hV+2)/2}} & \textrm{$h$ odd.}
\end{array}
\right.
\ee
It is worth stressing why $A'_{h,V}$ is the appropriate
coefficient: indeed it forces the graph to have at least one vertex
(as $A'_{h,0}=0$ in our definition), and has an appropriate symmetry
factor which accounts for the relative cyclic rotations of the link
pattern and the leaves in the tree.


So, in the case of $h$ odd, scaling the index $V$ in the sum by a
factor 2, we have
\be
\label{eq.47835}
Z_1(g)= \sum_V g^{2V} \frac{(2V(h+1))!}{(2V)! (hV+1)! (hV+2)!}
\ef,
\ee
from which we have the asymptotics
\be
\label{eq.asymphodd}
Z_1(g) \sim
\sum_V
\left( g \frac{(h+1)^{h+1}}{(h/2)^{h}} 
\right)^{2V}\!\! V^{-4}
\ef,
\ee
allowing to obtain the critical value of the coupling
\be
\label{eq.gcrit}
g_c(h)= 2^{-h} \frac{h^h}{(h+1)^{h+1}}
\ef,
\ee
and the universal exponent $-4$, which is in agreement with the KPZ
prediction. Indeed, we have just found that the string susceptibility
$\gamma$ of the random-graph spanning-tree model, 
defined through the relation
\be
Z \sim 
\sum_n 
x^n
n^{-3+\gamma}
\ef,
\ee
must be $\gamma=-1$. On the other side,
the central charge of the model in flat 2-dimensional space 
is $c=-2$. Finally, the KPZ relation consistently predicts
\be
\gamma = \frac{c-1- \sqrt{(25-c)(1-c)}}{12}
\ef.
\ee
Furthermore, if we recall that spanning forests emerge as a limit 
$q \to 0$ of the Potts model,
we also find that the value of the string susceptibility is in agreement
with the formula of Eynard and Bonnet \cite{eynbon}, valid for the
Potts Model in the range $q \in [0,4]$, that, in the parametrization
$q=2-2 \cos( \pi \nu)$ reads
\be
\gamma=1-\frac{2}{1 \pm \nu}
\ef.
\ee
From the ratio of two consecutive summands in (\ref{eq.47835}), we
deduce that the expression for $Z_1(g)$ is a generalized
hypergeometric function of variable $(g/g_c)^2$. In particular, for
the case of cubic planar lattices (shifting summation index from $V$
to $V-1$) we have the function
\be
\label{eq.Zsptreescubic}
Z_1(g) =
\frac{1}{12 g^2}
\left(
{}_2F_1 \left( -\smfrac{3}{4}, -\smfrac{1}{4}; 2; 2^6 g^2 \right)
-1 \right)
\ef.
\ee
In a similar way we can handle the case of $h$ even.
We have the formula
\be
\label{eq.47835bis}
Z_1(g)= \sum_V g^{V} \frac{(V(h+1))!}
{V! (\smfrac{1}{2} hV+1)! (\smfrac{1}{2} hV+2)!}
\ef,
\ee
from which we have the asymptotics
\begin{align}
\label{eq.asympheven}
Z_1(g) \sim \sum_V
\left( g \frac{(h+1)^{h+1}}{(h/2)^{h}} 
\right)^{V}\!\!\! V^{-4}
\ef,
\end{align}
which again gives the universal exponent $-4$, and the equation
(\ref{eq.gcrit}) for the critical value of the coupling.
From the ratio of two consecutive summands in (\ref{eq.47835bis}) we
deduce that 
we deal with a generalized
hypergeometric function of variable $g/g_c$. In particular, for the case
of coordination 4 (also in this case shifting summation index from $V$
to $V-1$) we have the function
\be
\label{eq.Zsptreesh2}
Z_1(g) =
\frac{1}{6 g}
\left(
{}_2F_1 \left( -\smfrac{2}{3}, -\smfrac{1}{3}; 2; 3^3 g \right) -1 \right)
\ef.
\ee
As we are interested to the limit of large volume, it is instructive
to analyze the approximate expressions 
(\ref{eq.asymphodd}) and (\ref{eq.asympheven}). Parametrizing 
$g/g_c = e^{-\epsilon}$ if $h$ is odd, and $g/g_c = e^{-2 \epsilon}$
if $h$ is even, and determining through Stirling expansion the overall
constants in (\ref{eq.asymphodd}) and (\ref{eq.asympheven}),
we get the unique expression
\begin{align}
t Z_1(g) 
&\sim 
t
\,
\kappa(h)
\sum_{n \geq 1} e^{-2 \epsilon n} n^{-4}
\ef;
&
\kappa(h)
&=
\left\{
\begin{array}{ll}
\displaystyle{\frac{\sqrt{h+1}}{2 \pi h^4}}
&
\textrm{$h$ odd,} \\
\rule{0pt}{19pt}%
\displaystyle{2 \frac{\sqrt{h+1}}{2 \pi h^4}}
&
\textrm{$h$ even.} \\
\end{array}
\right.
\end{align}
(The even/odd feature comes from the fact that, in the even case, the
summands have a double density).
The sum gives a polylogarithmic function, whose series expansion in
$\epsilon$ is
\be
\mathrm{Li}_4(e^{-2 \epsilon}) =
\zeta(4) - 2 \epsilon \, \zeta(3) + \cdots 
+ \frac{4}{3} \epsilon^3 \ln \epsilon
+ \cdots
\ef,
\ee
of which we highlighted the only term in the series which is not a
pure monomial in $\epsilon$, but involves the non-analyticity $\ln
\epsilon$ (this non-analiticity arising, of course, in addition to the
fact that the series is not convergent if $\epsilon < 0$). We
shall interpret the exact values of the terms in the series as being
affected by the approximation of the summands for small $n$, however this general feature
(that the first logarithmic term comes with a power $\epsilon^3$), as
it must be driven by the large-volume limit, is
well captured by the approximation, and even the corresponding
numerical coefficient $4/3$ should be exact. Indeed, the (more painful) series
expansion of the exact expressions 
(\ref{eq.Zsptreescubic}) and (\ref{eq.Zsptreesh2})
gives, for $h=1$,
\begin{align}
t \, \kappa(h)^{-1} Z_1(g) &=
\left( \frac{8192}{315} - \frac{16 \sqrt{2} \pi}{3} \right)
+ 
\left( \frac{14848}{315} - \frac{32 \sqrt{2} \pi}{3} \right)
\epsilon
+
\cdots
+
\frac{4}{3} \epsilon^3 \ln \epsilon
+ \cdots
\ef;
\end{align}
and for $h=2$,
\begin{align}
t \, \kappa(h)^{-1} Z_1(g) &=
\left( \frac{729}{5} - 24 \sqrt{3} \pi \right)
+ 
\left( \frac{648}{5} - 24 \sqrt{3} \pi \right)
\epsilon
+
\cdots
+
\frac{4}{3} \epsilon^3 \ln \epsilon
+ \cdots
\ef.
\end{align}
So we conclude that the leading non-analytic contribution to $Z_1$ 
(in powers of $\epsilon$) is given by
\be
\label{eq.epsF1}
t Z_1(g)
=
\left( \frac{t}{\epsilon} \right) \,
\left[
\left(
\textrm{$\mathcal{O}(\epsilon)$, analytic}
\right)
+
\kappa(h)
\frac{4}{3} \epsilon^4 
\ln \epsilon
\left(
1 + \mathcal{O}(\epsilon)
\right)
\right]
\ef.
\ee

\section{Perturbative expansion: two components}
\label{sec.pertF2}

Our goal is to construct a perturbative expansion for the partition
function (\ref{eq.Zfor}), at every order in $t$, using the
combinatorial results of section \ref{sec.precomb}.  The case of
spanning trees described in section \ref{sec.perttrees} was specially
simple.  Also the case of $2$-forests is slightly special, for a
number of concerns, while from the third order on we can give a
general recipe (this is done in section~\ref{sec.pertHO}).

Consider forests consisting of two trees, named as $F=(T_1, T_2)$. We have
\be
Z_2(g)= \frac{1}{2} \sum_{T_1, T_2} g^{V(T_1) + V(T_2)} 
\sum_{G \succ F} 
\frac{1}{|\aut(G \diagup F)|}
\ef,
\ee
where summation over $G$ is over connected planar graphs which contain
the two trees $T_{1,2}$, and no other vertices beside the ones in the
trees. The factor $\frac{1}{2}$ is due to the symmetry under exchange of the
two trees. Shrink the two trees $T_{1,2}$ to single vertices
$v_{1,2}$, and call $\deg(v)$ the degree of
vertex $v$. In order to have an integer number of vertices $N_{1,2}$
per tree, as $N_{1,2} = \frac{1}{h} ( \deg(v_{1,2}) - 2)$,
it must be $\deg(v_{1,2}) \equiv 2$ modulo $h$.

Now we can deal directly with the graph $G'=G \diagup F$, at the cost of a
combinatorial factor
for the number of ``reconstructions'' of the trees.
The 
edges of $G'$ are either bridges or
arcs\footnote{This is said following the terminology
of page~\pageref{page.bridgesarcs}, i.e.~an edge is an \emph{arc} if it
connects points on the same component, and a \emph{bridge} if it connects
points on distinct components.}.
For $\ell \geq 1$, consider the set $S_{\ell}$ of 
graphs $G'$ with $\ell$ bridges.
On any vertex, in the cyclic ordering, 
two subsequent bridge terminations are divided by a (possibly empty)
link pattern configuration.
The exact combinatorial factor is
\be
\label{eq.combfact1}
\frac{1}{\ell} A_{h,N_1} A_{h,N_2}
\ef,
\ee
plus the constraints that $\ell$, $N_1$ and $N_2$ are positive
integers. The symmetry reasonings leading to this result are as
follows. If we can fix a ``canonical'' leg among the ones incident on
$v_{1,2}$, we have a factor $A_{h,N}$, instead of the $A'_{h,N}$ that
one should combine with a summation on all the equivalent legs. A
candidate canonical leg is a leg of a bridge edge. However, we have
$\ell$ of them, ``equivalent'' because of the cyclic symmetry of the
definition, so we must sum over the marking of all the equivalent
bridge edges, and divide accordingly by~$\ell$.

Thus, we have a generating function (in some parameters $z_{1,2}^{-1}$, for
reasons explained in a moment)
\be
\frac{1}{\ell}
\left(
\frac{C(z_1^{-2})}{z_1}
\frac{C(z_2^{-2})}{z_2}
\right)^{\ell}
\ef,
\ee
and $Z_2$ is recovered from this
through the formal
substitution $z_{1,2}^{-(hN+2)} \to A_{h,N}$, and 
$z_{1,2}^{-N'} \to 0$ if $N' \not\equiv 2$ modulo $h$.

Remark the appearence of a recurrent combination
\be
\label{eq.qz}
q(z) :=
\frac{C(z^{-2})}{z}
=
\frac{1}{2} \Big( z-\sqrt{z^2-4} \Big)
\ee
(this is the expression for the resolvent of the Gaussian theory),
having the useful property
\be
\label{eq.qq2}
q(z)^2 = z q(z) - 1
\ef.
\ee
Summing over $\ell$ we are left with the function
\be
\label{eq.Qprimo}
\begin{split}
Q'(z_1, z_2)
&=
\sum_{\ell \geq 1} \frac{1}{\ell} 
(q(z_1)q(z_2))^{\ell}
=
- \ln (1 - q(z_1)q(z_2))
\ef.
\end{split}
\ee
%
Recalling that the function $A_h(z)$ is such that 
\be
\label{eq.firstoint}
\oint \frac{\dx{z}}{2 \pi i z}\; z^2 
\big( A_h(g z^{h}) - 1 \big)
z^{-(hn+2)} = A_{h,n} (1 - \delta_{n,0}) g^n
\ef,
\ee
we can use contour integration to implement the formal substitution above.
Define
\be
\label{eq.misa}
a(z)=
z \big( A_h(g z^{h}) - 1 \big)
\ef,
\ee
then we have
\be
\label{eq.Z2goback}
\begin{split}
Z_2(g)
& = 
\frac{1}{2}
\oint 
\!\! \prod_{j=1,2}
\frac{\dx{z_j}\,a(z_j)}{2 \pi i}
Q'(z_{1}, z_{2})
\ef.
\end{split}
\ee

\noindent
We would now need to determine the leading non-analytic term in the
limit $g \nearrow g_c$, i.e.~the two-component analogue of equation
(\ref{eq.epsF1}). For some reasons (namely, the cyclic symmetry of
Feynman diagrams constituted of two vertices and $\ell$ edges in
parallel, which leads to logarithmic series instead of geometric
ones), this calculation turns out to be a subtle variant of our
approach for forests with $n \geq 3$ components, analysed in the
following sections, and is discussed in full detail only in 
section~\ref{sec.feyneval}.

Here, and up to the end of the section, we will make a digression, in
which we use a different method, w.r.t.~the one of contour integration
in complex plane, used in the rest of the paper. This tool succeedes
in determining the nature of the leading non-analytic behaviour, but
fails to provide the exact value of the corresponding numerical
coefficient.


We will not enter too much in the necessary mathematical background of
the method (that we could call of ``L\'evy calculus''), however such a
pedagogical introduction should appear in a future companion paper,
where a number of other statistical results are derived in this way.

We go back to the understanding of $Z_2$ as a sum over the number
$\ell$ of bridges, and, given $\ell$, we have $2 \ell$ link pattern
configurations intertwined with the bridge legs, cyclically along the
two vertices. So we have
\be
\begin{split}
Z_2(g)
& = \sum_{\ell, N_1, N_2 \geq 1}
\frac{1}{2 \ell}
A_{h, N_1} A_{h, N_2}
\; g^{N_1 + N_2}
\\
& \qquad \times
\sum_{\nu_1, \ldots, \nu_{2 \ell} \geq 0}
\delta_{h N_1 + 2, 2(\nu_1 + \cdots + \nu_{\ell}) + \ell}
\;
\delta_{h N_2 + 2, 2(\nu_{\ell + 1} + \cdots + \nu_{2\ell}) + \ell}
\,
\prod_{\alpha=1}^{2 \ell}
C_{\nu_{\alpha}}
\ef.
\end{split}
\ee
We can easily manipulate a number of overall factors. Define the
rescaled Catalan numbers as
\be
\hat{C}_n = 2^{-2n-1} C_n
\ef,
\ee
such that the corresponding generating function has radius of
convergence 1, and indeed these coefficients are normalized,
$\sum_{n \geq 0} \hat{C}_n = 1$. Define also the rescaled coefficients
$\hat{A}_{h,N}$ as
\be
\hat{A}_{h,N}
=
\big( 2^h g_c(h) \big)^n
A_{h,N}
\ee
so that also their generating function has radius of
convergence 1 (it is however normalized to $1+\frac{1}{h}$).
Through these coefficients, the leading exponential factors are
clearly highlighted
\be
\begin{split}
Z_2(g)
& = 2^4 \sum_{\ell, N_1, N_2 \geq 1}
\frac{1}{2 \ell}
\hat{A}_{h, N_1} \hat{A}_{h, N_2}
\; \left( \frac{g}{g_c} \right)^{N_1 + N_2}
\\
& \qquad \times
\sum_{\nu_1, \ldots, \nu_{2 \ell} \geq 0}
\delta_{h N_1 + 2, 2 (\nu_1 + \cdots + \nu_{\ell}) + \ell}
\;
\delta_{h N_2 + 2, 2 (\nu_{\ell + 1} + \cdots + \nu_{2\ell}) + \ell}
\,
\prod_{\alpha=1}^{2 \ell}
\hat{C}_{\nu_{\alpha}}
\ef.
\end{split}
\ee
The parameter $\ell$ is the only variable which entangles the
contributions coming from the two trees, so for a given value of
$\ell$ we can investigate separately each term
\begin{align}
\label{eq.Z2ff}
Z_2(g)
& = \sum_{\ell \geq 1}
\frac{1}{2 \ell}
f_{\ell}(g)^2
\ef;
\\
f_{\ell}(g)
& = \sum_{N \geq 1}
\hat{A}_{h, N} 
\; \left( \frac{g}{g_c} \right)^{N}
\sum_{\nu_1, \ldots, \nu_{\ell} \geq 0}
\delta_{h N + 2, 2 (\nu_1 + \cdots + \nu_{\ell}) + \ell}
\,
\prod_{\alpha=1}^{\ell}
\hat{C}_{\nu_{\alpha}}
\ef.
\end{align}
The quantity $f_{\ell}$ can be easily calculated exactly for $\ell=1$
and $h=1$, with a method analogous to the one of section
\ref{sec.perttrees}. The result is
\be
\begin{split}
f_1(g_c e^{- \epsilon})
& = 
\sum_{\nu \geq 1}
\hat{A}_{1, 2 \nu - 1}
\hat{C}_{\nu}
e^{- \epsilon (2 \nu - 1)}
=
e^{\epsilon}
\left(
1 - {}_2F_1(-\smfrac{1}{4},\smfrac{1}{4};2;e^{-2 \epsilon})
\right)
\\
& =
\bigg(1 - \frac{32 \sqrt{2}}{15\pi} \bigg)
+ \cdots
- \frac{1}{4 \pi} \epsilon^2 \ln \epsilon
+ \cdots
\ef,
\end{split}
\ee
where we highlighted the leading contribution, and the leading
non-analytic contribution. So, the leading 
contribution to $f_{1}^2$ is of order 1, while 
the leading non-analytic contribution comes from the cross product,
and is of order $\epsilon^2 \ln \epsilon$. If we factor out a term
$\left( t/\epsilon \right)^2$ in $t^2 Z_2$, we have a combination of the
kind
\be
\label{eq.combination}
\left( \frac{t}{\epsilon} \right)^2
\left( A (\epsilon^2 + \cdots) + B (\epsilon^4 + \cdots) \ln \epsilon
\right)
\ef,
\ee
analogous in form to (\ref{eq.epsF1}), with the difference that the
analytic term starts from a higher order.

We will now study the function $f_{\ell}(g)$ for arbitrary values of
$\ell$, and in particular in an approximation valid for large values.
At this point, we make use of the core of L\'evy calculus, i.e.~the
generalized Central Limit Theorem for variables sampled from a
heavy-tailed distribution (see for example \cite[sec.~3.7]{levy1}).
Defining ``the'' L\'evy distribution\footnote{Some authors name this
distribution after L\'evy, while others use the name of L\'evy
distributions for the broader family of alpha-stable distributions.}
\be
\levy_c(x) = \sqrt{\frac{c}{2 \pi}} e^{-\frac{c}{2 x}}
x^{-\frac{3}{2}}
\ef,
\ee
which is a special representative of the L\'evy family of alpha-stable
distributions, for $\alpha=\frac{1}{2}$ and $\beta=1$, with
pseudo-variance $c$. We have
\be
\label{eq.89764756875}
\sum_{\nu_1, \ldots, \nu_{\ell} \geq 0}
\delta_{N, \nu_1 + \cdots + \nu_{\ell}}
\,
\prod_{\alpha=1}^{\ell}
\hat{C}_{\nu_{\alpha}}
\simeq
\levy_{\frac{\ell^2}{2}}(N + \mathcal{O}(\ell))
\ef,
\ee
and, in particular, typical values of $N$ are of order $\ell^2$.  The
reason for this is that rescaled Catalan numbers have the asymptotics
$\hat{C}_n \sim \frac{1}{\sqrt{4 \pi}} n^{-\frac{3}{2}}$, so that the
tail of the distribution on the integers $\hat{C}_n$ matches with
$\levy_{\frac{1}{2}}(x)$, and the sum of $N$ independent variables
with parameter $\alpha$ and pseudo-variances $c_i$ follows an alpha-stable
distribution with parameter $\alpha$, and pseudo-variance $c$
satisfying $c^{\alpha} = \sum_i c_i^{\alpha}$. A more subtle analysis
(namely, a second order in Stirling approximation for the Catalan numbers)
would allow to state in (\ref{eq.89764756875})
\be
\levy_{\frac{\ell^2}{2}}(N + \mathcal{O}(\ell))
=
\levy_{\frac{\ell^2}{2}} 
\big( N - \smfrac{7}{12} \ell + \mathcal{O}(1) \big)
\ef,
\ee
so, in order to control the errors deriving from the linear part in
$\ell$, we will write $N - b \ell$ as argument of the L\'evy distribution.

At this point an issue of factors $2$ coming from an even/odd feature
emerges. The point is the constraint that $2( \nu_1 + \cdots +
\nu_{\ell}) + \ell \equiv 2$ modulo $h$. If $h$ is odd, this happens
approximatively with a flat probability $1/h$, when $N$ is large,
regardless of $\ell$.  If $h$ is even, this can never happen if $\ell$
is odd, and happens approximatively with a flat probability $2/h$,
when $N$ is large and $\ell$ is even. To be definite, we will consider
the case of odd $h$, although it would not be difficult to treat also
the other case.  We get
\be
\sum_{\nu_1, \ldots, \nu_{\ell} \geq 0}
\delta_{h N + 2, 2 (\nu_1 + \cdots + \nu_{\ell}) + \ell}
\,
\prod_{\alpha=1}^{\ell}
\hat{C}_{\nu_{\alpha}}
\simeq
\frac{1}{h}
\levy_{\frac{\ell^2}{2}}
\big( \smfrac{h}{2} (N - b \ell) \big)
\ef.
\ee
Parametrizing $g = g_c e^{-\epsilon}$, and using the asymptotic
behaviour of $\hat{A}_{h,N}$ (as deduced from (\ref{eq.adefSt})), one has
\be
f_{\ell}(g) \simeq 
\sum_{N \geq 1}
\frac{ \sqrt{h+1} \, \ell }{\pi h^3}
N^{-3}
\exp \left(
-\frac{\ell^2}{2 h N}
- \frac{\epsilon h}{2} (N+b \ell)
\right)
\ef.
\ee
We see how the integrand has a good scaling for $N \sim \ell^2$
and $\epsilon \sim \ell^{-2}$, with the correction term in $b$
subleading (of order $\frac{1}{\ell}$). We will neglect it from now
on.

Approximating the sum with an integral (again, legitimate up to
corrections of relative order $\frac{1}{\ell}$), and using the known
formula
\be
\int_0^{\infty} \frac{\dx{x}}{x^{3}} 
\exp \left(-a x -\smfrac{b}{4x} \right) = 
\frac{8a}{b} K_2(\sqrt{ab})
\ef,
\ee
where $K_2(x)$ is the Bessel $K$ function of index 2, we get
\be
f_{\ell}(g) \simeq 
\frac{ \sqrt{h+1} \, \ell }{\pi h^3}
\frac{2 \epsilon h^2}{\ell^2}
K_2 (\sqrt{\epsilon} \ell)
=
\epsilon
\frac{2 \sqrt{h+1}}{\pi h \ell}
K_2 (\sqrt{\epsilon} \ell)
\ef.
\ee
Substituting into the expression (\ref{eq.Z2ff}) for $Z_2$, we have
\be
Z_2(g)
\simeq
\epsilon^2
\frac{2 (h+1)}{(\pi h)^2}
\sum_{\ell \geq 1}
\frac{1}{\ell^3}
K_2 (\sqrt{\epsilon} \ell)^2
\ef.
\ee
In this case, the leading non-analitic behaviour in $\epsilon$ must be
deduced from the characteristics of Bessel function $K_2$ near the
origin (as the contribution at large $\ell$ is suppressed both by the
algebraic prefactor, and by the behaviour of $K_2$ itself).
We have
\be
K_2(2z) =
\left( \frac{1}{2 z^2} + \cdots \right)
- \ln z \left( \frac{z^2}{2} + \cdots \right)
\ef,
\ee
where the dots stand for further terms in a series expansion. So, the
leading term in the square goes like $\frac{1}{4 z^4}$, while the
leading non-analitic term comes from the cross product, and goes like
$-z^2 \ln z \, K_2(2 z)$. Using also the formula
\be
\int_a^{\infty} \frac{\dx{x}}{x} K_2(x) = \frac{K_1(a)}{a}
= \frac{1}{a^2} (1 + \mathcal{O}(a^2 \ln a))
\ef,
\ee
we get
\be
\label{eq.1326456454}
\sum_{\ell \geq 1}
\frac{1}{\ell^3}
K_2 (\sqrt{\epsilon} \ell)^2
=
\left(
4 \zeta(7)
\epsilon^{-2}
+ \cdots
\right)
- \frac{1}{2} \ln \epsilon
\left(
\frac{1}{4} 
+ \cdots
\right)
\ef,
\ee
and finally
\be
\label{eq.epsF2}
t^2 Z_2(g)
=
\left( \frac{t}{\epsilon} \right)^2
\left[
\left(
\textrm{$\mathcal{O}(\epsilon^2)$, analytic}
\right)
-
\frac{
(h+1)}{(2 \pi h)^2}
\epsilon^4 \ln \epsilon
(1 + \mathcal{O}(\epsilon) )
\right]
\ef.
\ee
The fact that the leading non-analytic behaviour is dominated by the
cross product (thus breaking the symmetry between the two trees), and
that the leading contribution to the sum (\ref{eq.1326456454}) comes
from small values of $\ell$, is a hint towards the fact that the
generating function $Z_2$, in its limit of large graphs, is dominated
by forests in which one of the two trees is much larger than the
other. We will come back to this point in sections
\ref{sec.feyneval} and~\ref{sec.nonana}.

Bessel functions do not appear anymore in the paper, so there should
not be confusion with an unrelated quantity, defined in the following,
for which we use the letter~$K$.

\section{Perturbative expansion: higher orders}
\label{sec.pertHO}

Here we show how to construct a perturbative expansion for the
partition function (\ref{eq.Zfor}), at every order in $t$, using the
combinatorial results of section \ref{sec.precomb}.  The cases of
spanning trees and spanning forests with two components, described
respectively in sections \ref{sec.perttrees} and \ref{sec.pertF2},
were special under certain aspects, while from the third order on we
can give a general recipe, being a slight modification of the
technique of Cauchy integral explained at the beginning of section
\ref{sec.pertF2} (conversely, we do not use anymore the tools of
``L\'evy calculus'' adopted in the remaining part of that section).

At the light of the `hardness' of the derivation for 2-forests, in
comparison with the one for trees, the topic of the present section
could seem an ambitious task. However, a more careful comparison of
formulas (\ref{eq.epsF1}) and (\ref{eq.epsF2}) leads to more
optimistic expectations. Indeed, from these partial results, it looks
like the series $Z(g) = \sum_n t^n Z_n(g)$, for $g=g_c e^{-\epsilon}$
near to the critical value, can be written as a double series in the
variables $t/\epsilon$ and $\epsilon$. In formulas (\ref{eq.epsF1})
and (\ref{eq.epsF2}) we highlighted both the \emph{tout-court} leading
behaviour, and the leading behaviour among the terms containing some
non-analiticity for $\epsilon \to 0$ (namely, a factor $\ln \epsilon$). Of
course, the more the latter is subleading, the more a
refined control on the result is necessary, in order to extract a
sufficient number of terms in a full expansion. Conversely, if the
latter is leading, a first-order perturbative analysis should be
sufficient. From the analysis of the cases with $n$ components, $n$
being 1 or 2, it can be conjectured that the disturbing analytic
series starts with $\epsilon^n$, while the
first non-analytic term occurs at order $\epsilon^4 \ln \epsilon$ (in
agreement with the value of the string susceptibility, and the fact
that $|V| \sim 1/\epsilon$). As
a result, if such a conjecture did hold for all values of $n$, for 
$n \geq 4$ we would be in the easier situation 
in which first-order perturbative analysis suffices.


So we start the analysis of the combinatorics for forests 
$F=(T_1, \ldots, T_n)$, with $n \geq 3$, being spanning on a graph 
$G \succ F$.
Shrink the trees to vertices $v_1$, \ldots, $v_n$. The graph obtained
so far has been called $G'=G \diagup F$. Imagine $G'$ as drawn on the Riemann
sphere (i.e., on the plane, plus a point at infinity). Given an
arc, two regions on
the Riemann sphere are naturally identified. Say that the arc is
\emph{contractible} if at least
one of the two regions contains only arcs.
For the remaining edges,
given two edges connecting the same pair of vertices (they could be 
two bridges or two non-contractible arcs), again two regions on
the Riemann sphere are naturally identified. Say that the
edges are \emph{multiple} if one of the two regions contains only 
arcs.\footnote{Remark that, if
  $n \geq 3$, at most one of the two regions has this property.
  This is one of the reasons why the case $n=2$ is special.}
Call $G''(G')$ the graph in which contractible arcs are
removed, and $G'''(G')$ the one in which both contractible arcs are
removed, and multiple edges are replaced by single edges,
by shrinking out the regions containing only arcs. This procedure is
depicted in figure \ref{fig_765465}.
By construction,
$G'''$ does not contain neither contractible arcs, nor consecutive multiple
edges. It is easily seen that, at fixed $n$, the number of planar
graphs with these characteristics is finite, and the number of
edges is at most $3n-6$.\footnote{This is a consequence of Euler
  formula (\ref{eq.Euler}) for connected planar graphs, i.e.~with
  $\genus=0$ and $K=1$, $V+F-E=2$.
Furthermore, the absence of contractible arcs
and of multiple edges implies that all the faces have at least 3 sides,
so that $2E \geq 3F$, equality holding for triangulations. Solving
w.r.t.~$F$ gives the desired relation
\be E \leq 3V-6 \ef. \ee}
In a sense, the graph $G'''$ describes the relevant part of the
adjacence structure among components of the forest. We have a `bridge'
edge between two vertices if the corresponding trees are adjacent in
the full graph, and a non-contractible loop incident on a vertex if
the corresponding tree $T$ has edges in $G \smallsetminus T$ incident
on vertices of $T$ with both terminations, and which leave some other
components on each of the two sides.

The removal described above corresponds to a resummation analogous to
the one done for 2-forests, in determining the combination
(\ref{eq.Qprimo}).
Consider the generating function for
graphs $G'$ with $n$ vertices
\be
\label{eq.Z'n}
Z'_n = \sum_{G'} 
\frac{1}{|\aut(G')|}
\prod_{i\in V(G')} x_i^{\deg(i)}
\ef,
\ee
which we will put in relation with the desired perturbative partition function
\be
\label{eq.ZnZ'n}
Z_n = \sum_{G'} 
\frac{1}{|\aut(G')|}
\prod_{i\in V(G')} \!\!A_{h, \frac{\deg(i)-2}{h} }
\, g^{\frac{\deg(i)-2}{h}}
\ef.
\ee
In equation (\ref{eq.Z'n}),
the contribution of an edge between vertices $i$ and $j$ is $x_i x_j$,
so we can equivalently write
\be
Z'_n= \sum_{G'} 
\frac{1}{|\aut(G')|}
\prod_{(ij) \in E(G')} 
(x_i x_j)
\ef.
\ee
Given a vertex with a certain number of terminations of bridges and
non-contractible arcs, each interval between these
terminations can be occupied by contractible arcs in an arbitrary
link pattern configuration. Thus, we can ``dress'' the edge
terminations with the substitution
\be
x_i \to x_i C(x_i^2)
\ef,
\ee
and then restrict the sum to all graphs $G''$ which do not contain any
contractible arc,
\be
\label{eq.4696325}
\begin{split}
Z'_n
&=
\sum_{G''}
\frac{1}{|\aut(G'')|}
\prod_{i=1}^n (x_i C(x_i^2))^{\deg(i)}
\\
&=
\sum_{G''}
\frac{1}{|\aut(G'')|}
\;
\prod_{\substack{e=(i,j)\\ \in E(G')}} 
( x_i C(x_i^2) x_j C(x_j^2) )
\ef.
\end{split}
\ee
Then, multiple edges can be resummed. For each $\ell$-uple of
multiple edges between vertices $i$ and $j$, we have a contribution
$(x_i C(x_i^2) x_j C(x_j^2))^{\ell}$. As each edge in $G'''$ can be
originated by shrinking an arbitrary number $\ell \geq 1$ of multiple
edges, we can replace them with dressed single edges through the
substitution
\be
\label{eq.87554}
x_i C(x_i^2) x_j C(x_j^2)
\to
\frac{x_i C(x_i^2) x_j C(x_j^2)}{1 - x_i C(x_i^2) x_j C(x_j^2)}
\ef,
\ee
to be performed in the second formulation of quantity~(\ref{eq.4696325}),
together with a further restriction in the sum to graphs $G'''$ which do
not have multiple edges
\be
Z'_n
=
\sum_{G'''}
\frac{1}{|\aut(G''')|}
\prod_{\substack{e=(i,j)\\ \in E(G''')}} 
\frac{x_i C(x_i^2) x_j C(x_j^2)}{1 - x_i C(x_i^2) x_j C(x_j^2)}
\ef.
\label{eq.Z'feyn}
\ee
We claim that the formal substitution
rule $x_j^{hN+2} \to A_{h,N} g^N$ if $N$ is a positive integer, and
$x_j^{N'} \to 0$ otherwise, applied to $Z'_n$, gives $Z_n$. The only
delicate point is the symmetry factor pertinent to vertices, i.e.~we
need to prove that a canonical way exists for marking a single leg
incident to any vertex, without any other symmetry factors. 
Label arbitrarily the vertices with integers $1$, \ldots, $n$. This
accounts for the already included obvious factor $|\aut(G''')|^{-1}$.
Remark that, as $G'''$ is connected, each vertex must have at least one
incident bridge, and that a canonical spanning tree on $G'''$ is
easily constructed, e.g.~the depth-first tree starting from the `root'
vertex $1$. 

\begin{figure}[t]
\begin{center}
\setlength{\unitlength}{25pt}
\begin{picture}(13.2,5)
\put(0,0.3){\includegraphics[scale=1.1, bb=0 300 300 400,
    clip=true]{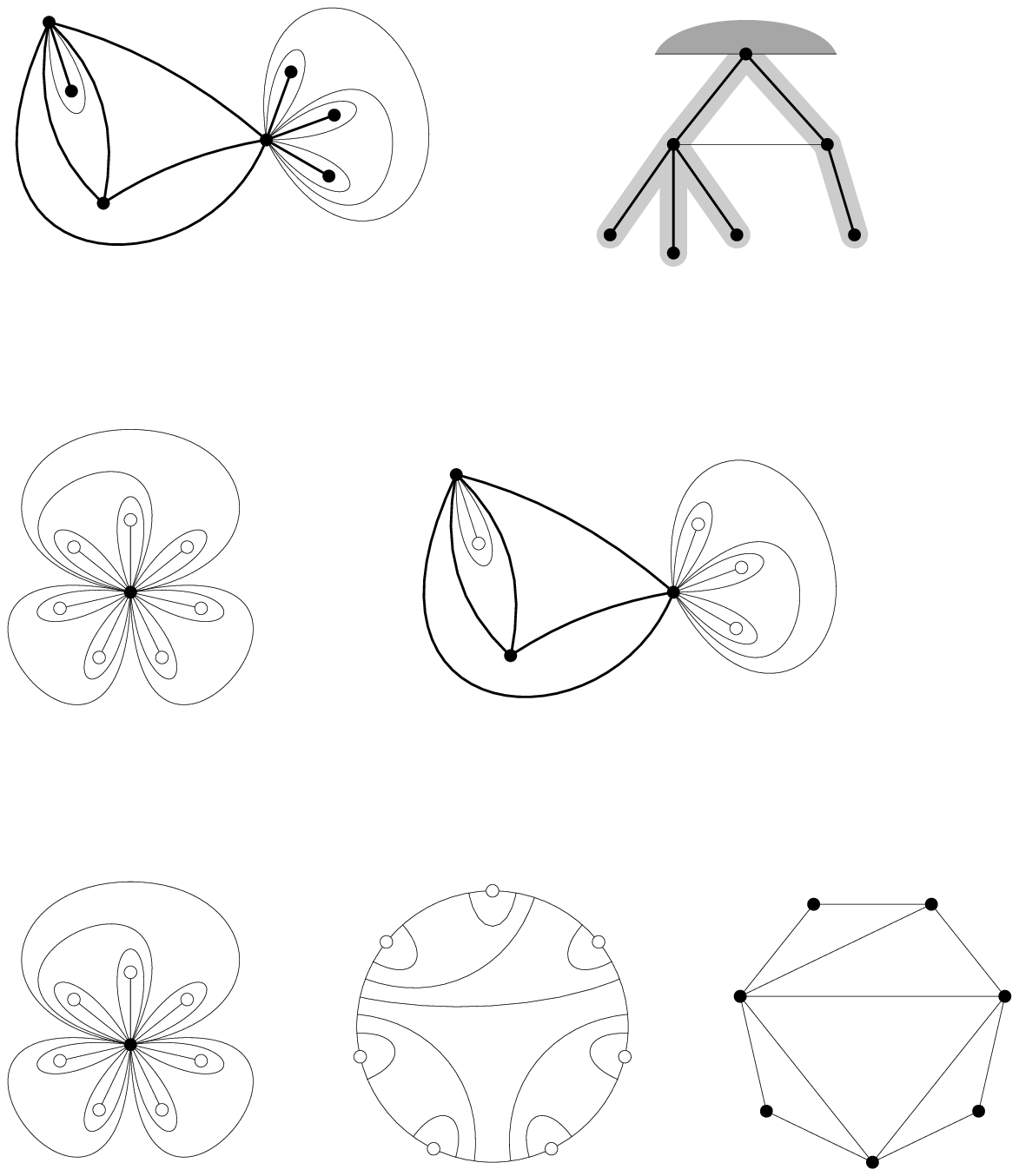}}
\put(1.8,1.2){$\scriptstyle{1}$}
\put(3.92,1.85){$\scriptstyle{2}$}
\put(4.2,3.5){$\scriptstyle{3}$}
\put(1,2.6){$\scriptstyle{4}$}
\put(0.9,4.1){$\scriptstyle{5}$}
\put(5.05,1.65){$\scriptstyle{6}$}
\put(5.1,2.75){$\scriptstyle{7}$}

\put(10.4,3.3){$\scriptstyle{1}$}
\put(9.5,2.3){$\scriptstyle{2}$}
\put(8.7,1.05){$\scriptstyle{3}$}
\put(12.2,0.7){$\scriptstyle{4}$}
\put(11.8,2.0){$\scriptstyle{5}$}
\put(9.55,0.55){$\scriptstyle{6}$}
\put(10.55,0.6){$\scriptstyle{7}$}
\end{picture}
\mycaption{\label{fig.DFtree}On the left, a typical diagram, with labeled
  vertices. Bridges are drawn in bold. On the right, the corresponding
  depth-first tree, on the graph in which non-consecutive multiple
  edges are replaced by a single edge, and arcs are dropped. Remark
  that, in building the depth-first tree, the planar embedding of the
  diagram is discarded, and a new one is induced by the vertex
  labeling. In particular, the loop surrounding the tree is the
  contour of the light-gray region.}
\end{center}
\end{figure}

Note that this depth-first tree has a natural orientation towards the
root, so that any non-root vertex has a single outgoing edge, oriented
towards his neighbour with smaller index.  Also, this tree has a
natural planar embedding in $\HH$ (with the root vertex on $\de \HH$),
and thus a natural oriented loop which encircles it (see figure
\ref{fig.DFtree} for an example).

For any non-root vertex $i$, having vertex $j$ as a neighbour with
minimum index, we will choose as ``canonical'' leg one of the bridge
edges between $i$ and $j$. We have in general $L \geq 1$ such edges.
As $n \geq 3$, at least one among $i$ and $j$ has other neighbours, so
that the $L$-uple of bridges is univocally splitted into $m$
$\ell$-uples of multiple edges, and in any $\ell$-uple a ``first''
edge in a cyclic ordering is identified (this is a difference with the
$n=2$ case, where no ``first leg'' can be defined). We will use as a canonical
leg one of these first legs in a $\ell$-uple, so we are left with $m
\geq 1$ choices. We have to prove that a canonical choice among these
$m$ can be
performed. At this point, the loop surrounding the tree plays a
role. Indeed, the chosen pair $(ij)$ is traversed by the loop, from
$i$ to $j$, exactly once, and thus a single ``next'' vertex $k$, such
that $(jk)$ follows $(ij)$ along the loop, is identified. Because
of planarity, this vertex must be contained in exactly one of the
regions identified by the $m$ multiple-edges, and thus it also
identifies a canonical choice for a leg, as was to be proven.
Now that we have many reference points, for the root vertex any
reasonable choice makes the game. For example, one can use the other
termination of the edge containing the canonical leg of the neighbour
with smaller index. This completes the discussion of the involved
symmetry factors.

So, we can use the technique described in formula
(\ref{eq.firstoint}), in order to extract $Z_n$ from $Z'_n$, i.e.
\begin{figure}[t]
\begin{center}
\setlength{\unitlength}{25pt}
\begin{picture}(9.,10.75)
\put(0.3,0){\includegraphics[scale=0.85, bb=0 410 235 720,
    clip=true]{figs.eps}}
\end{picture}
\mycaption{\label{fig_765465}On the top, a typical configuration of graph
  $G'$ with $n=4$ vertices. Contractible arcs are drawn as thin edges, and
  dotted lines collect multiple edges. On the bottom, the resulting
  graph $G'''(G')$.}
\end{center}
\end{figure}
\begin{figure}[ht]
\begin{center}
\setlength{\unitlength}{20pt}
\begin{picture}(19,8)(0,0)
\put(0,0){\includegraphics[scale=0.6, bb=250 670 950 950,
    clip=true]{figs.eps}}
\put(1,7.){$\displaystyle{\frac{1}{3!}}$}
\put(5,7.){$\displaystyle{\frac{1}{2}}$}
\put(4.8,3){$\displaystyle{\frac{1}{3!}}$}
\put(8.8,7){$\displaystyle{\frac{1}{2}}$}
\put(12.4,7){$\displaystyle{\frac{1}{2}}$}
\put(16.0,7){$\displaystyle{\frac{1}{2}}$}
\put(8.4,3){$\displaystyle{\frac{1}{8}}$}
\put(12.0,3){$\displaystyle{\frac{1}{8}}$}
\put(15.4,3){$\displaystyle{\frac{1}{4!}}$}
\end{picture}
\mycaption{\label{fig.feyn}Diagrams of the perturbative expansion for
  three-component (top-left) and
four-component forests. Linear combinations, leading to the appearance
of $\Qp$ propagators, are chosen in order to minimize the total number
of diagrams. The coefficients, deriving from this combination of
symmetry factors, are indicated next to each diagrams. Remark that the
first non-contractible arc appears at order 3, in the right-most
diagram, while the first non-consecutive multiple edges appear in some
of the diagrams at order~4.}
\end{center}
\end{figure}
\be
\label{eq.7642654}
Z_n=\prod_{j=1}^n 
\left( \oint \frac{\dx{z_j} a(z_j)}
{2 \pi i} \right) 
Z'_n(\{ z_j^{-1} \})
\ef.
\ee
The quantity in (\ref{eq.87554}), when expressed in terms of inverse
parameters $\{ z_j^{-1} \}$, corresponds to
\be
\label{eq.defQm}
\frac{
 \frac{C(z_1^{-2})}{z_1}
 \frac{C(z_2^{-2})}{z_2} }
{ 1-
 \frac{C(z_1^{-2})}{z_1}
 \frac{C(z_2^{-2})}{z_2} }
=
\frac{q(z_1) q(z_2)}{1- q(z_1) q(z_2)}
=:
\Qm(z_1,z_2)
\ef,
\ee
where $q(z)$ has been defined in (\ref{eq.qz}), and $\Qm(z_1,z_2)$ is a
useful combination.

For future convenience, we also define
\be
\Qp(z_1,z_2)
:=
1+ \Qm(z_1,z_2)
=
\frac{1}{1- q(z_1) q(z_2)}
\ef,
\ee
and remark that these quantites can be restated as
\begin{align}
\label{eq.Qpmii}
\Qpm(z_i, z_i)
&=
\frac{1}{2} \left(
\frac{ z_i }{\sqrt{z_i^2-4}}
\pm 1 \right)
\ef;
\\
\label{eq.Qpmij}
\Qpm(z_i, z_j)
&=
\frac{1}{2} \left(
\frac{ \sqrt{z_i^2-4} - 
\sqrt{z_{\jind}^2-4} }{z_i - z_j}
\pm 1 \right)
\ef;
\end{align}
and in particular\footnote{Equation 
  (\ref{eq.Qpmii}) comes from (\ref{eq.Qpmij}) through a l'H\^opital
  limit, while (\ref{eq.Qpmij}), which coincides with (\ref{eq.Qmij})
  trivially, is related to the definition (\ref{eq.defQm}) through the
  use of property (\ref{eq.qq2}). Namely, calling $q=q(z)$ and
  $q'=q(z')$, by verifying that
\[
\frac{q q'}{1- q q'} = - \frac{q - q'}{z - z'}
\ef.
\]
}
\begin{align}
\label{eq.Qmii}
\Qm(z_i, z_i)
&=
\frac{q(z_i)}{\sqrt{z_i^2 - 4}}
\ef;
\\
\label{eq.Qmij}
\Qm(z_i, z_j)
&=
- \frac{q(z_i) - q(z_j)}{z_i - z_j}
\ef.
\end{align}
In equation (\ref{eq.7642654}), expanded with (\ref{eq.Z'feyn}), we
can interchange integrations and summation over the graphs.  Then,
each term of the sum can be interpreted as a (coordinate-space)
``Feynman diagram'': indeed, we have the appropriate symmetry factor,
one integration per vertex, with a proper ``measure'' factor $a(z_j)$,
and a product of ``propagators'', or Green functions, $\Qm(z_i, z_j)$
corresponding to the edges of the diagram.
\be
Z_{n}=\sum_{G'''}
\frac{1}{|\aut(G''')|}
\oint \!\! \prod_{j=1}^{n}
\frac{\dx{z_j}\,a(z_j)}{2 \pi i}
\!\!\!
\prod_{\substack{e=(i,j)\\ \in E(G''')}} 
\!\!\!\!
\Qm(z_i, z_j)
\ef.
\ee
In section \ref{sec.feyneval}
we will deal with the problem of understanding the appropriate
contours for such an integral, and we will prove that a valid contour
exists if and only if $g \leq g_c(h)$.
The integrals for the third and fouth order are
\begin{align}
\begin{split}
Z_3 
&=
\oint \prod_{j=1}^{3} 
\frac{\dx{z_j}\,a(z_j)}{2 \pi i}
\;
\frac{1}{6} \Qm_{12} \Qm_{23} (\Qm_{31} + 3 \Qp_{22})
\ef;
\end{split}
\\
\begin{split}
Z_4
&=
\oint \prod_{j=1}^{4} 
\frac{\dx{z_j}\,a(z_j)}{2 \pi i}
\,
\bigg[
\frac{1}{2} \Qm_{12} \Qm_{23} \Qm_{34}
\bigg(
\frac{1}{4}
\Qm_{14} \Qp_{13} \Qp_{24}
+
\Qp_{22} \Qp_{33} \Qp_{23}
\bigg)
\\
& \; 
+
\frac{1}{2} \Qm_{12} \Qm_{13} \Qm_{14}
\bigg(
{\Qp_{11}}^2 \Qm_{12}
+ 
\frac{1}{3} {\Qp_{11}}^3
+
\Qp_{11} \Qp_{12} \Qm_{23}
+
\frac{1}{12}
\Qp_{23} \Qp_{34}
(3 \Qm_{12} + \Qm_{24} )
\bigg)
\bigg]
\ef;
\end{split}
\end{align}
where short notation $\Qpm_{ij}$ stands for $\Qpm(x_i, x_j)$, 
and in many cases the contributions of more diagrams are
collected together, with the combination $\Qp_{ij} = 1 + \Qm_{ij}$.
The related diagrams are shown in 
figure \ref{fig.feyn}, where dashed lines denote propagators $\Qp_{ij}$.

Here we remark that this approach in some sense ``captures'' the
degrees of freedom of the perturbative theory: we sum over the
contribution of an infinite number of degrees of freedom (as we are
summing over graphs with an arbitrary number of vertices), and we end
up with a finite diagrammatics, i.e.~a finite sum of
finite-dimensional integrals. Remark in particular that, as expected,
the diagrams in our family are all planar.

A detailed discussion on how to practically estimate the result of
these integrations appears in section~\ref{sec.feyneval}.

We remark that most of the techniques outlined in this paper for
the spanning-forest model immediately generalize to variants of the
model, in which more general weights are chosen for the trees
\be
Z_{\textrm{gen.}}= \sum_{G} \frac{N^{-2 \genus(G)}}{|\aut(G)|}
g^{|V(G)|}
\sum_{F \prec G} \frac{|\aut(G)|}{|\aut(G \diagup F)|}
\prod_{T_{\alpha} \in F} (t \, w(T_{\alpha}))
\ef;
\ee
here the weights $w(T)$ for the single components (i.e.~the trees)
of the forest only depend on the structure of the tree (and not on the
embedding into $G$). Our case is $w(T)=1$. Other interesting cases are
$w(T)=|T|$ (\emph{rooted forests}, cfr.~below);
the generalization $w(T)=|T|^{\nu}$; the case 
$w(T)=\sum_{v \in V(T)} \xi_{\deg_T(v)}$, which is the generating function
for the distribution of vertex coordinations. In all these cases, the
weight $w(T)$ reflects into a modification of the generating function
$a(z)$ in (\ref{eq.misa}), while all the formulas in which $a(z)$
appears implicitly are valid in the generic case. In particular, the
case $w(T)=|T|^{\nu}$, for various values of $\nu$, explores all the
asymptotic behaviours of $a(z)$, ``shifting'' the spanning-tree
string susceptibility $\gamma=-1$ to $\gamma+\nu$. This is absolutely
trivial in the case of trees, as $|T| = |V(G)|$, but has substancial
consequences on the measure for spanning forests at $t > 0$. In
particular, for $\nu=1$ we have a model of \emph{rooted} spanning
forests, a considerably easier variant of spanning forests,
corresponding to a massive perturbation of the pure graph-Laplacian
implicit in Kirchhoff Matrix-Tree theorem.

\section{Evaluation of diagrams}
\label{sec.feyneval}

In the previous section we stated some combinatorial quantities in
terms of contour integrals, which formally express certain
convolutions of generating functions. In this section we
discuss the analytic prescriptions on how to perform these
integrations, and give a general technique to reduce our integrals
to integrals of real-valued functions over real intervals.

We give a number of examples, in increasing order of difficulty, with
the aim of gradually introducing a set of tools, more extensively used
in the following sections. For some of these examples, also the
results are of direct interest.

The easiest example of this method is
still given by Catalan numbers. Suppose one wants to deduce formula 
(\ref{eq.catal}) from the generating function
(\ref{eq.catalgf}). Then, one can perform a 
contour integration
\be
C_n = 
\oint \frac{\dx{z}}{2 \pi i z}\; \frac{1-\sqrt{1-4 z}}{2 z} z^{-n}
\ef.
\ee
It is legitimate to
deform the contour of integration, up to encircle the whole
plane except for the cut $[\smfrac{1}{4}, +\infty]$:
\[
\setlength{\unitlength}{30pt}
\begin{picture}(9,4)
\put(0,0){\includegraphics[scale=1, bb=240 445 510 570, clip=true]{figs.eps}}
\put(2.75,1.65){$\frac{1}{4}$}
\put(7.35,1.55){$\frac{1}{4}$}
\end{picture}
\]
For $n$ sufficiently large ($n > -\smfrac{1}{2}$ suffices, so for any
integer $n$), the integral on the large circle vanishes, and we are
left with the integral on the two sides of the cut.
For $z=x\pm i \epsilon$, with $x$ real larger than $\smfrac{1}{4}$ and 
$\epsilon$ real positive infinitesimal, one has 
$\sqrt{1 - 4 (x\pm i \epsilon)} = \mp i |\sqrt{4x-1}| +
\mathcal{O}(\epsilon)$, and one can write
\be
C_n = 
\int_{\frac{1}{4}}^{+\infty} \frac{\dx{x}}{2 \pi}\; 
\frac{\sqrt{4x - 1}}{x^{n+2}}
\ef,
\ee
which gives indeed Catalan numbers.

A next case, of intermediate difficulty between the previous example
and the general-diagram case, is the case of spanning trees on cubic
lattices. Suppose we want to deduce the result of equation
(\ref{eq.Zsptreescubic}) with no use of the explicit coefficients
$C_{n}$ and $A'_{1,n}$, but only through the generating functions
$C(x)$ and $A'_1(x)$. Start consider the case of generic $h$. Call $n$
the index for the tree size, and $m$ the one for the number of edges,
then the spanning-tree partition function reads
\be
Z_1(g)=\sum_{n,m} g^n A'_{h,n} C_m \delta_{2m, hn+2}
\ef.
\ee
In generating function,
\begin{gather}
x^2 A'_{h} (g x^h) = \sum_n g^n A'_{h,n} x^{hn+2}
\ef,
\\
C(x^{-2}) = \sum_m C_m x^{-2m}
\ef,
\end{gather}
and contour integration can be used in order to reproduce the delta
constraint
\be
Z_1(g)=
\oint \frac{\dx{z}}{2 \pi i z}\; 
z^2 A'_{h} (g z^h) C(z^{-2})
\ef.
\ee
For the case $h=1$, using equations (\ref{eq.catalgf}) and
(\ref{eq.a'1}), we have
\begin{gather}
z^2 A'_{1} (g z) =
\frac{-1+6 g z - 6 (gz)^2 + (1-4 gz)^{\frac{3}{2}}}
{12 g^2}
\ef,
\\
\frac{1}{z} 
C(z^{-2}) = 
\frac{z - \sqrt{z^2-4}}{2}
\ef.
\end{gather}
Now it is clear which prescription we should use for the contour
integral: we should encircle the origin, being 
in the radius of convergence for both series
$C(x)$ and $A'_1(x)$, thus $2 < |z| < 1/(4g)$. As
expected, no contour can be found when $g > g_c=1/8$. In 
terms of Cauchy integration,
this prescription states that the contour should
encircle the cut going from $-2$ to $2$, and leave
outside the cut going from $1/(4g)$ to infinity. If we deform the
path in order to have contribution only from the $[-2,2]$ cut
discontinuity, as in
\[
\setlength{\unitlength}{30pt}
\begin{picture}(5.66,3.5)(0,0.3)
\put(0,0){\includegraphics[scale=1, bb=770 445 940 570, clip=true]{figs.eps}}
\put(1.2,1.5){$-2$}
\put(4.12,1.5){$2$}
\put(4.7,1.5){$\frac{1}{4g}$}
\end{picture}
\]
we have
\be
\label{eq.Z1h1}
Z_1(g)= \frac{1}{12 g^2} 
\int_{-2}^2 \frac{\dx{x}}{2 \pi}\; 
\sqrt{4-x^2}
\Big(-1+6 g x - 6 (gx)^2 + (1-4 gx)^{\frac{3}{2}}
\Big)
\ef.
\ee
It can be seen that this is the appropriate expression.
Indeed, if we
expand the generating function $A'_1(x)$, keep only even orders
(because of the parity of the other factor in the integrand),
and use the formula
\be
\int_{-2}^2 \frac{\dx{x}}{2 \pi}\; 
x^{2n} \sqrt{4-x^2} 
=C_n
\ee
we obtain the series coefficients (\ref{eq.47835}). 

What would have happened if we wanted to evaluate the integral by the
change of variables which makes $A'_1(x)$ an analytic function? The
new variable would have been $x'=x(1-g x)$. For $g$ infinitesimal, the
cut $[-2,2]$ moves infinitesimally to the two new values
\begin{align}
x_+ &= \frac{1 - \sqrt{1 - 8g}}{2g} = 2 + 4 g + \cdots
\ef,
\\
x_- &= \frac{1 - \sqrt{1 + 8g}}{2g} = -2 + 4 g + \cdots
\ef,
\end{align}
and a new cut, to be left outside the contour
of integration, appears, between the values
\begin{align}
x'_+ &= \frac{1 + \sqrt{1 - 8g}}{2g} = \frac{1}{g} - 2 -4g + \cdots
\ef,
\\
x'_- &= \frac{1 + \sqrt{1 + 8g}}{2g} = \frac{1}{g} + 2 -4g + \cdots
\ef.
\end{align}
Above the critical value $g_c=1/8$, the two solutions $x_+$ and $x'_+$,
instead of being radially ordered, are complex conjugate, and there is
no contour in which the convergence of the generating function is
assured.

This phenomenology occurs also at higher values of $h$, with a slight
even/odd difference. Indeed, one can study the counter-images of the
cut $z \in [-2,2]$ under the map $z=x(1-g x^h)$. The value $z=0$ has
$h+1$ counter-images $x=\{ 0 \} \cup \{ g^{-\frac{1}{h}} e^{ \frac{2
\pi i \ell}{h} } \}_{0 \leq \ell < h}$, which `mark' the $h+1$
counter-images of the cut if $g < g_c$ (where the critical value $g_c$
is defined in (\ref{eq.gcrit})). However, at $g=g_c$, two real
counter-images of $z=2$, for the central cut and the cut with label
$\ell = 0$, collapse, at the point $x_{\star} = 2 \frac{h+1}{h}$. If
$h$ is even, by symmetry this also happens to two real counter-images
of $z=-2$, the ones for the central cut and for $\ell = h/2$. An
example for $h=7$, at the critical coupling, is illustrated in
figure~\ref{fig.taglini}.

\begin{figure}[!tb]
\begin{center}
\setlength{\unitlength}{26.4pt}
\begin{picture}(9.6,9.6)(0,0)
\put(0,0){\includegraphics[scale=1.5, bb=770 225 940 395, clip=true]{figs.eps}}
\put(6.7,7){$g_c^{\frac{1}{h}} e^{\frac{2 \pi i \ell}{h}}$}
\put(6.4,4.25){$\frac{2(h+1)}{h}$}
\end{picture}
\mycaption{\label{fig.taglini}The counter-images of the interval
  $[-2,2]$ under the map $z=x(1-g_c x^h)$, in the complex plane for
  $x$. Here $h=7$. The bullets correspond to the counter-images of the
  origin. The dashed circle denotes the largest disk centered in the
  origin which does not intersect any of the external cuts, and is
  thus the radius of convergence for the integrand corresponding to a
  typical diagram in the Feynman expansion. A valid contour must
  encircle the internal cut, staying within this disk.}
\end{center}
\end{figure}

Another check of the validity of equation (\ref{eq.Z1h1}), and a
different approach to the result of equation (\ref{eq.epsF1}), comes
from deducing that, for $h=1$, 
\[
Z_1(g) = 
(\textrm{$\mathcal{O}(1)$, analytic})
+
\kappa(1)
\frac{4}{3} \epsilon^3
\ln \epsilon
\left(
1 + \mathcal{O}(\epsilon)
\right)
\ef.
\]
This is equivalent to state that
\be
\frac{\partial^3}{\partial \epsilon^3}
Z_1(g) = 
8 \kappa(1)
\ln \epsilon
+ \mathcal{O}(1)
\ef.
\ee
Indeed, in (\ref{eq.Z1h1}) the parameter $\epsilon$ occurs only in the
``measure'' $A'_1(g z)$, and one has, for $z \sim 2$ and $g=g_c e^{-\epsilon}$,
\be
\begin{split}
\frac{\partial^3}{\partial \epsilon^3}
(z^2 A'_1(g z))
& \sim
-\frac{3}{8}
\frac{(4gz)^3}{12 g^2}
(1-4 g z)^{-\frac{3}{2}}
+
\textrm{more reg.~terms}
\\
& =
-2
(1-4 g z)^{-\frac{3}{2}}
+
\textrm{more reg.~terms.}
\end{split}
\ee
The semicircle factor $\sqrt{4-z^2}$, near the singularity $z=2$, is
approximated by $2 \sqrt{2-z}$. Then, integrating in some window
$[2-a,2]$ near the singularity, we get
\be
Z_1 \sim
-\frac{2}{\pi}
\int_{2-a}^2 \dx{z}
\;
\sqrt{2-z}
\;
\Big(
1-\frac{z}{2} e^{-\epsilon}
\Big)^{-\frac{3}{2}}
=
\frac{4 \sqrt{2}}{\pi} \ln \epsilon + \mathcal{O}(1)
=
8 \kappa(1) \ln \epsilon + \mathcal{O}(1)
\ef,
\ee
as was to be shown.

Now we repeat the procedure above in the case of $Z_2(g)$, in order to
determine the coefficient of the leading non-analytic term, at least
for $h=1$, which was lacking in section \ref{sec.pertF2}. 
From the L\'evy calculus arguments of that section, we have an
expansion of the form
\[
Z_2(g) = 
(\textrm{$\mathcal{O}(1)$, analytic})
+
c \, \epsilon^2
\ln \epsilon
\left(
1 + \mathcal{O}(\epsilon)
\right)
\ef,
\]
with $c$ an unknown numerical constant.
This is equivalent to state that
\be
\frac{1}{2}
\frac{\partial^2}{\partial \epsilon^2}
Z_2(g) = 
c
\ln \epsilon
+ \mathcal{O}(1)
\ef.
\ee
We have to go back to the expression (\ref{eq.Z2goback}). 
This expression is symmetric under exchange $z_1 \leftrightarrow z_2$,
and the parameter $\epsilon$ occurs only in the
``one-body measures'' $a(z_1)$ and $a(z_2)$. So we can write
\be
\label{eq.Z2restart}
\begin{split}
\frac{1}{2}
\frac{\partial^2}{\partial \epsilon^2}
Z_2(g)
& = 
\frac{1}{2}
\frac{\partial^2}{\partial \epsilon^2}
\frac{1}{2}
\oint 
\frac{\dx{z_1}\,a(z_1)}{2 \pi i}
\oint 
\frac{\dx{z_2}\,a(z_2)}{2 \pi i}
\; Q'(z_{1}, z_{2})
\\
& = 
\frac{1}{2}
\oint 
\frac{\dx{z_1}}{2 \pi i}
\left( \frac{\partial^2}{\partial \epsilon^2} a(z_1) \right)
\oint 
\frac{\dx{z_2}\,a(z_2)}{2 \pi i}
\; Q'(z_{1}, z_{2})
\\
& \quad +
\frac{1}{2}
\oint 
\frac{\dx{z_1}}{2 \pi i}
\left( \frac{\partial}{\partial \epsilon} a(z_1) \right)
\oint 
\frac{\dx{z_2}}{2 \pi i}
\left( \frac{\partial}{\partial \epsilon} a(z_1) \right)
\; Q'(z_{1}, z_{2})
\\
& =: \mathcal{A} + \mathcal{B}
\ef.
\end{split}
\ee
We will prove in the following that the second summand,
$\mathcal{B}$, does not contribute to the leading non-analytic part of
$Z_2$. We start by calculating the leading expression corresponding to
the first integral, $\mathcal{A}$. It is useful to integrate by parts w.r.t.~variable
$z_1$, in order to get a restatement which avoids logarithmic
functions.  This is affordable in principle also for arbitrary $h$,
because the function $a(z)$, of which little is known, has however a
simple primitive (cfr.~equation (\ref{eq.aa'})), and would give
\be
\mathcal{A}
=
\frac{1}{2}
\oint 
\frac{\dx{z_1}}{2 \pi i}
\left( \frac{\partial^2}{\partial \epsilon^2} (z_1^2 A'_h(g z_1^h))
\right)
\frac{1}{\sqrt{z_1^2-4}}
\oint 
\frac{\dx{z_2}\,a(z_2)}{2 \pi i}
\; \Qm(z_{1}, z_{2})
\ef.
\ee
The two derivatives acting on variable $z_1$ produce a singularity
$\sim 1/z$ near $z_1=2$, up to more regular terms, so that, neglecting
other more regular terms, we can replace $z_1 \to 2$ in all
non-singular expressions, and in particular in $\Qm$, thus factorizing
the two integrals. As discussed later (in equation (\ref{eq.defKh})),
the remaining integral in $z_2$ is regular for $\epsilon \to 0$, and
named $K_h$. So we can write
\be
\label{eq.backhereZ2}
\mathcal{A}
=
\frac{K_h}{2}
\oint
\frac{\dx{z}}{2 \pi i}
\left( \frac{\partial^2}{\partial \epsilon^2} (z^2 A'_h(g z^h))
\right)
\frac{1}{\sqrt{z^2-4}}
+
\textrm{more reg.~terms.}
\ee
Again we can deform the contour in order to take contribution from the
cut, and get
\be
\mathcal{A}
=
\frac{K_h}{2 \pi}
\int_{-2}^{2} 
\dx{z}
\left( \frac{\partial^2}{\partial \epsilon^2} (z^2 A'_h(g z^h))
\right)
\frac{1}{\sqrt{4 - z^2}}
+
\textrm{more reg.~terms.}
\ee
and one has, for $h=1$, $z \sim 2$ and $g=g_c e^{-\epsilon}$,
\be
\begin{split}
\frac{\partial^2}{\partial \epsilon^2}
(z^2 A'_1(g z))
& \sim
\frac{3}{4}
\frac{(4gz)^2}{12 g^2}
(1-4 g z)^{-\frac{1}{2}}
+
\textrm{more reg.~terms}
\\
& =
4 (1-4 g z)^{-\frac{1}{2}}
+
\textrm{more reg.~terms,}
\end{split}
\ee
so that
\be
\label{eq.Z2h1}
\mathcal{A} \sim
\frac{K_1}{\pi}
\int_{2-a}^2 \dx{z}
\;
\frac{1}{\sqrt{2-z}}
\;
\Big(
1-\frac{z}{2} e^{-\epsilon}
\Big)^{-\frac{1}{2}}
=
-\frac{\sqrt{2}}{\pi} K_1 \ln \epsilon + \mathcal{O}(1)
\ef.
\ee
More generally, for arbitrary $h$, going back to
(\ref{eq.backhereZ2}), we have
\be
\frac{\partial^2}{\partial \epsilon^2} (z^2 A'_h(g z^h))
=
\frac{1}{h^2}
\left(
z^2
\frac{\partial^2}{\partial z^2}
-
3 z
\frac{\partial}{\partial z}
+ 4
\right)
(z^2 A'_h(g z^h))
\ef.
\ee
Higher derivatives lead to stronger singularities, so the leading
order is due to the second-order derivative, and we can write
\be
\frac{\partial^2}{\partial \epsilon^2} (z^2 A'_h(g z^h))
\simeq
\frac{1}{h^2}
z^2
\frac{\partial^2}{\partial z^2}
(z^2 A'_h(g z^h))
=
\frac{1}{h^2}
z^2
\frac{\partial}{\partial z}
a(z)
\ef.
\ee
Having replaced derivatives w.r.t.~$\epsilon$ with derivatives
w.r.t.~$z$ allows us to perform the $g$-dependent change of variables
which simplifies the measure.  In changing variables from $z$ to $x$,
the Jacobian in the remaining derivative and in the measure simplify,
and we have in general
\be
\oint
\frac{\dx{z}}{2 \pi i}
\left(
\frac{\partial}{\partial z}
a(z)
\right)
F(z)
=
\oint
\frac{\dx{x}}{2 \pi i}
\left(
\frac{\partial}{\partial x}
g x^{h+1}
\right)
F \big( x(1+g x^h) \big)
\ef,
\ee
while in our specific case
\be
\label{eq.backhereZ2b}
\mathcal{A}
\simeq
\frac{K_h}{2h^2}
\oint
\frac{\dx{x}}{2 \pi i}
\big( 
(h+1) g x^h
\big)
\frac{\big( x (1+g x^h) \big)^2 }
{\sqrt{\big( x (1+g x^h) \big)^2 - 4}}
\ef.
\ee
We can now reduce to an integral over the cut discontinuity due to the
square root at denominator.
Suppose that $h$ is odd, so that we have a single point of singularity
(instead of two symmetric ones). This singularity occurs at the
rightmost extremum of the central cut (cfr.~figure \ref{fig.taglini}),
where two counter-images $x_*^{(1,2)}$ of $z=2$ coincide for 
$g \nearrow g_c$. Denote with index $1$ the most internal solution,
and with $x_* = 2 \frac{h+1}{h}$ the limit for $g=g_c$.

As we are only concerned with the
leading singularity, which is due to the square root at denominator
$\sqrt{\big( x (1+g x^h) \big) - 2}$, in all other factors we can
replace the value of $g$ with $g_c$, and of $x$ with $x_*$, up to less
singular terms, and write
\be
\label{eq.backhereZ2c}
\mathcal{A}
\simeq
\frac{K_h}{2 \pi h^2}
\int^{x_*^{(1)}}
\dx{x}
\cdot 1 \cdot
\frac{\big( 2 \big)^2 }
{\sqrt{4} \sqrt{\big( x (1+g x^h) \big) -2}}
=
\frac{K_h}{\pi h^2}
\int^{x_*^{(1)}}
\dx{x}
\frac{1}
{\sqrt{x (1+g x^h) -2}}
\ef.
\ee
Furthermore, in the polynomial of order $h+1$, 
$x (1+g x^h) -2$, we should highlight the two main roots
$x = x_*^{(1,2)}$, and for the rest we can replace the other roots
with their values for $g=g_c$. Using the fact
\be
\lim_{g \to g_c}
\frac{x (1+g x^h) -2}{(x-x_*^{(1)}) (x-x_*^{(2)})}
=
\lim_{g \to g_c}
\left.
\frac{1}{2}
\frac{\partial^2}{\partial x^2}
\big( x (1+g x^h) -2 \big)
\right|_{x=x_*}
=
-\frac{h^2}{4(h+1)}
\ef,
\ee
we can write
\be
\label{eq.backhereZ2d}
\mathcal{A}
\simeq
\frac{K_h}{\pi h^2}
\int_{0}
\dx{x}
\frac{2 \sqrt{h+1}}{h}
\frac{1}
{\sqrt{x (\delta x_*-x)}}
=
\frac{2 K_h \sqrt{h+1}}{\pi h^3}
\big( -\ln \delta x_* + \mathcal{O}(1) \big)
\ef,
\ee
with the shortcut $\delta x_* := x_*^{(2)}-x_*^{(1)}$.
As we have
\be
x_*^{(2)}-x_* \sim x_*-x_*^{(1)} \sim 
\sqrt{\frac{2}{h(h+1)}} \sqrt{\epsilon}
\ef,
\ee
we have
\be
-\ln \delta x_*
=
- \frac{1}{2} \ln \epsilon + \mathcal{O}(1)
\ef,
\ee
so that we get
\be
\label{eq.Z2genh}
\mathcal{A}
=
- \frac{K_h \sqrt{h+1}}{\pi h^3}
\big( \ln \epsilon + \mathcal{O}(1) \big)
\ef.
\ee
This is in agreement with the special case $h=1$ of
equation~(\ref{eq.Z2h1}).

For what concerns the integral $\mathcal{B}$, we have that the two
one-body measures produce a singularity $\sim 1/\sqrt{z}$ for any of
the variables $z_{1,2} \to 2$. Beside this, there is only a two-body
interaction factor $Q'(z_1,z_2)$, which is regular unless \emph{both}
varables approach $2$, in which case, for $z_1, z_2$ along the cut
$[-2,2]$ and near $2$, it behaves as 
\be
Q'(z_1,z_2) \sim \frac{1}{2} \ln \big( \max (2-z_1,2-z_2) \big)
\ee
so that, even taking directly $\epsilon=0$, we get a finite result
after integrating near to the only potential singularity
\be
\mathcal{B}
\propto
\int_0^a \dx{z_1}
\int_{z_1}^a \dx{z_2}
\;
\frac{\ln z_2}{\sqrt{z_1 z_2}}
=
2 a (\ln a - 1)
\label{calb}
\ef.
\ee
By collecting (\ref{eq.Z2restart}),  (\ref{eq.Z2genh}) and (\ref{calb}) we get
\be
Z_2(g_c e^{-\epsilon})
=
\left[
( \textrm{$\mathcal{O}(1)$ analytic, $\mathcal{O}(\epsilon^3)$} )
-
\frac{\sqrt{h+1} K_h}{\pi h^3}
\; \epsilon^2
\ln \epsilon
\right]\ef. \label{z2}
\ee
The integrals arising from the full perturbative expansion are
similar, although more involved, and, of course,
multi-dimensional. However, as from three components on the symmetry
factors are easier to handle, there is a small variety of fundamental
ingredients: only the one-body measure $a(z)$ and the two-body
interaction $\Qm(z,z')$ occur.
Consider a diagram $D$ with $n$ vertices and
edge-set $E(D)$. The integral is of the form
\be
\label{eq.ID}
\mathcal{I}(D) =
\prod_j \oint
\frac{\dx{z_j}\,a(z_j)}{2 \pi i}
\prod_{(ij) \in E(D)}
\Qm(z_i, z_j)
\ef.
\ee
Consider the change of variables $z_j = x_j (1-g x_j^h)$. The 
measure changes as
\begin{align}
\label{eq.dzaz}
\dx{z} \,
a(z)
&=
\dx{x}\;
gx^{h+1}(1-(h+1)g x^h)
\ef.
\end{align}
We will keep using $z_j$ as a shortcut to $x_j (1-g x_j^h)$, and $q_j$
as a shortcut of $q(z_j)$, when this does not cause confusion. Recall
that it is the expression $\sqrt{z_{\jind}^2 - 4}$ in $q(z_j)$ which is
discontinuous at the cut, and that, in the complex plane for $x$, the cut
to be encircled by the integration contour is the real segment
$[x_-(g), x_+(g)]$, with $|x_-|, |x_+| \leq 2 \frac{h+1}{h}$, and
$x_+ (g) \to 2 \frac{h+1}{h}$ for $g \to g_c$.

There are two `easy' contour prescriptions, in the complex planes for
$x_j$'s, which are valid for any value of $h \geq 1$, and $g \leq
g_c(h)$. One is the circle of radius $2 \frac{h+1}{h}$, so that one
could use an ``angular'' parametrization 
$x_j = 2 \frac{h+1}{h} e^{i \theta_j}$, with $\theta_j \in [0, 2\pi]$.
A second one is to consider a ``cut'' integral, i.e.~integrate along
the sides of a rectangle, centered at the origin, and of half-sides 
$2 \frac{h+1}{h} \times \delta$, in a limit $\delta \to 0$. In this
limit only the cut discontinuity survives. 

In section \ref{sec.nonana} we concentrate on the angular
parametrization, when trying to extract the leading behaviour for 
$g \to g_c$. Here we discuss briefly the way in which the cut integral
should be performed. Concentrate on a given variable $x_i$.
Through the expressions (\ref{eq.Qmii}) and (\ref{eq.Qmij}), the
product of $\Qm$ edge terms is, up to a prefactor not involving $z_i$,
a function of the form
\[
\bigg(
\frac{q_i}{\sqrt{z_i^2 - 4}}
\bigg)^{m_i}
\prod_{j \neq i}
\left( \frac{q_i - q_j}{z_i - z_j} \right)^{m_j}
\]
with $m$'s being non-negative integers. The polynomial in $q_i$ in the numerator
can be reduced to a linear function in $q_i$, of the form $P_0(z_i) +
P_1(z_i) q_i$, by iterated use of (\ref{eq.qq2}). If $m_i$ is even, only
the square root cut discontinuity caused by the $\sqrt{z_i^2 - 4}$
term in $q_i$ gives contribution to the integral, so that we have
\be
\begin{split}
&
\oint \prod_{j \neq i} \dx{z_j} \cdots
\oint \frac{\dx{z_i}}{2 \pi i} a(z_i)
\frac{
P_0(z_i) + P_1(z_i) q_i
}{
(4 - z_i^2)^{\frac{m_i}{2}} 
\prod_{j \neq i} (z_i - z_j)^{m_j}
}
\\
& \qquad 
=
\oint \prod_{j \neq i} \dx{z_j} \cdots
\int_{-2}^2 \frac{\dx{z_i}}{2 \pi} 
\frac{a(z_i)}{ \sqrt{4 - z_i^2} }
\frac{P_1(z_i) 
}{
(4 - z_i^2)^{\frac{m_i-2}{2}} 
\prod_{j \neq i} (z_i - z_j)^{m_j}
}
\ef.
\end{split}
\ee
If instead $m_i$ is odd, a cut term is already included in the
denominator, so only the regular terms in the numerator contributes
\be
\begin{split}
&
\oint \prod_{j \neq i} \dx{z_j} \cdots
\oint \frac{\dx{z_i}}{2 \pi i} a(z_i)
\frac{
P_0(z_i) + P_1(z_i) q_i
}{
(4 - z_i^2)^{\frac{m_i}{2}} 
\prod_{j \neq i} (z_i - z_j)^{m_j}
}
\\
& \qquad 
=
\oint \prod_{j \neq i} \dx{z_j} \cdots
\int_{-2}^2 \frac{\dx{z_i}}{2 \pi} 
\frac{a(z_i)}{ \sqrt{4 - z_i^2} }
\frac{2 P_0(z_i) + 
z_i P_1(z_i) 
}{
(4 - z_i^2)^{\frac{m_i-1}{2}} 
\prod_{j \neq i} (z_i - z_j)^{m_j}
}
\ef.
\end{split}
\ee
In particular, a simple case of these integrals, in some variable
$z_i$, is when $i$ is adjacent to only one vertex $j$, through $\ell$
edges. In this case we have
\be
\label{eq.PhiL}
\begin{split}
\Phi^{(\ell)}(w; g) 
&= 
\oint \frac{\dx{z}}{2 \pi i} a(z)
\left( \frac{q(w) - q(z)}{z - w} \right)^\ell
\\
&=
\int_{-2}^2 \frac{\dx{z_i}}{2 \pi} 
a(z_i) \frac{ \sqrt{4 - z_i^2} }{(z - w)^\ell}
\sum_h \binom{\ell}{2h+1}
\left( q(w)-\frac{z}{2} \right)^{\ell-2h-1}
(z^2-4)^h
\ef.
\end{split}
\ee
The case $\ell=1$, in the specialization $h=1$ and $g=g_c$, gives
\be
\label{eq.Phi}
\begin{split}
\Phi^{(1)}(w; g_c) 
&= 
- \oint \frac{\dx{z}}{2 \pi i} a(z)
\frac{q(z) - q(w)}{z - w}
=
- \int_{-2}^2 \frac{\dx{z}}{2 \pi} 
a(z) \frac{ \sqrt{4 - z^2} }{z - w}
\\
&=
\frac{4 \sqrt{2}}{3 \pi} (2 - 3w)
+ (1 - (w-4) q(w))
- \frac{\sqrt{2}}{\pi} 
\frac{ 4 - w^2 }{ \sqrt{w+2} }
\ln \frac{\sqrt{w+2} + 2}{\sqrt{w+2} - 2}
\ef.
\end{split}
\ee
For $w$ near to the endpoint of the cut,
\be
\label{eq.Phi_ser}
\begin{split}
\Phi^{(1)}(2+x; g_c)
&=
\bigg(
3 - \frac{16 \sqrt{2}}{3 \pi}
\bigg)
- 2 \sqrt{x}
- \frac{1}{\pi} x \ln x
+ \frac{4 \sqrt{2}}{\pi} (2 \ln 2 - 1) \, x
+ \mathcal{O}(x^{\frac{3}{2}})
\ef.
\end{split}
\ee
A calculation similar to the one performed in (\ref{eq.Phi}) leads to
$\left. \frac{\dx{}}{\dx{\epsilon}} \Phi^{(1)}(w; g_c e^{- \epsilon})
\right|_{\epsilon = 0}$, which is a long expression that we do not
write here. However, we report the equivalent of (\ref{eq.Phi_ser})
\be
\label{eq.DPhi_ser}
\begin{split}
\left.
\frac{\dx{}}{\dx{\epsilon}}
\Phi^{(1)}(2+x; g_c e^{- \epsilon})
\right|_{\epsilon = 0}
&=
\frac{2 \sqrt{2}}{\pi}
\ln x
+
\left(
4 + 
\frac{4 \sqrt{2}}{3 \pi}
-
6 \sqrt{2} \ln 2
\right)
+ \mathcal{O}(x^{\frac{1}{2}})
\ef.
\end{split}
\ee
Collecting the two results of (\ref{eq.Phi_ser})
and (\ref{eq.DPhi_ser}) we finally obtain
\be
\label{eq.Phi_ser2}
\begin{split}
\Phi^{(1)}(2+x; g_c e^{- \epsilon})
&=
\bigg(
3 - \frac{16 \sqrt{2}}{3 \pi}
\bigg)
- 2 \sqrt{x}
- \frac{1}{\pi} x \ln x
+ \frac{4 \sqrt{2}}{\pi} (2 \ln 2 - 1) \, x
\\
& \quad +
\frac{2 \sqrt{2}}{\pi}
\epsilon \ln x
+
\left(
4 + 
\frac{4 \sqrt{2}}{3 \pi}
-
6 \sqrt{2} \ln 2
\right)
\epsilon 
+ \mathcal{O}(x^{\frac{1}{2}} \epsilon, x^{\frac{3}{2}})
\ef.
\end{split}
\ee
We thus see that the limit for $w \to 2$ of
$\Phi^{(1)}$ is finite. Curiously, it coincides with the one for 
$\Phi^{(2)}$, that is
\be
\label{eq.phi1phi2}
\Phi^{(1)}(2; g)
=
\Phi^{(2)}(2; g)
\ef,
\ee
as the difference in the integrand is given by a factor
(cfr.~the general expression (\ref{eq.PhiL}))
$
\binom{2}{1}
\frac{q(w)-z/2}
{w - z}
$,
which goes to 1 identically in $z$, for $w \to 2$.

\section{Non-analytic behaviour for $g \nearrow g_c$}
\label{sec.nonana}

In order to highlight the leading contribution to $\{ Z_n \}_{n \geq
3}$ in the large-volume limit $g \nearrow g_c$, we consider the series
expansion described in section \ref{sec.pertHO}, in ``angular
parametrization'', that is parametrizing the $x_j$'s and $g$ as
\begin{align}
x_j
&=
\frac{2(h+1)}{h} e^{i \theta_j}
\ef;
&
g 
&= 
g_c e^{-\epsilon}
\ef.
\end{align}
In particular, the recurrent combination $g x^h$ becomes
$\frac{1}{h+1} e^{-\epsilon + i h \theta}$. The various relevant
quantities become
\begin{align}
\label{eq.was118}
\frac{\dx{z}}{2 \pi i}
a(z)
&=
\frac{\dx{\theta}}{2 \pi}
\,
\frac{4(h+1)}{h^2}
\,
e^{-\epsilon + i (h+2) \theta}
(1-e^{-\epsilon + i h \theta})
=:
\frac{\dx{\theta}}{2 \pi} \mu(\theta)
\ef;
\\
\label{eq.was119}
z
&=
2 e^{i \theta} 
\,
(1+\smfrac{1}{h} (1- e^{-\epsilon + i h \theta}) )
\ef;%
\rule{0pt}{13pt}\raisebox{-6pt}{\rule{0pt}{13pt}}
\\
\label{eq.was120}
\sqrt{z^2-4}
&=
z  \sqrt{1-e^{-2 i \theta} (1+\smfrac{1}{h} 
 (1- e^{-\epsilon + i h \theta}) )^{-2} }
\ef;
\end{align}
We will also adopt the shortcut $f(\theta_i, \theta_j)$ for
$\Qm(z(\theta_i), z(\theta_j))$.

There are various potential sources of non-analiticities for $g
\nearrow g_c$, due to non-regularities of the integrand for $\theta_i
\to 0$ or $\theta_i - \theta_j \to 0$ for the various indices, when
$\epsilon$ approaches $0$.

A first singularity may come from the pole $z_i-z_j$ in the
denominators of $\Qpm(z_i,z_j)$. However, for generic values of $z_i$,
this is not a true singularity, as, for $z_i \to z_j$, also the
numerator vanishes with the same behaviour (this is what allowed us to
determine an expression for $\Qpm(z_i,z_i)$). A first source of true
non-analiticity comes, in the limit $\theta_j \to 0$,\footnote{Or
  $\theta_j \to \pi$, if $h$ is even, however we neglect this for
  simplicity. It would be easy to reintroduce certain factors 2
  overall at the end, in the case of even $h$, while the present
  treatment covers the case of $h$ odd.}
from combinations of the form
\be
\label{eq.675867}
1-e^{-\epsilon + i h \theta}
=
\epsilon - i h \theta + \smfrac{1}{2} h^2 \theta^2
+
\mathcal{O}(\epsilon^2, \epsilon \theta, \theta^3)
\ef.
\ee
Consistently with the fact that we drop higher orders in $\theta$,
we may adopt this approximation only in some small window 
$\theta \in [-\delta, \delta]$, $\delta \ll 1$.

We expanded at second order in $\theta$ because, in the integration,
only overall even monomials contribute. The measure
$\mu(\theta)$
gives at leading orders, besides a factor as in
(\ref{eq.675867}),
\begin{align}
\label{eq.675867b}
e^{-\epsilon + i (h+2) \theta}
& =
1+i (h+2) \theta 
+ \mathcal{O}(\epsilon, \theta^2)
\ef.
\end{align}
The product of (\ref{eq.675867}) and (\ref{eq.675867b}) gives
\be
\label{eq.measthet}
\epsilon 
- i h \theta
+ \frac{h(3h+4)}{2} \theta^2
+
\mathcal{O}(\epsilon^2, \epsilon \theta, \theta^3)
\ef.
\ee
We will see in a moment that terms odd in $\theta$ do not play any
role in this measure.

Indeed, a further source of non-analiticity for $\theta \to 0$ is the
combination $z-2$, appearing as a factor in the square roots inside
terms $\Qpm$. In this case, a stronger cancellation, also of the terms
linear in $\theta$, occurs.  We get for the combination in
(\ref{eq.was119}) 
\be
\label{eq.zappr}
z
=
2
+
\smfrac{2}{h}
\left(
\epsilon + \smfrac{h(h+1)}{2} \theta^2
\right)
+
\mathcal{O}(\epsilon^2, \epsilon \theta, \theta^3)
\ef,
\ee
and thus we can rewrite the expression in (\ref{eq.was120}) as
\be
\label{eq.cheso}
\frac{z^2-4}{z^2}
=
\smfrac{2}{h}
\left(
\epsilon + \smfrac{h(h+1)}{2} \theta^2
\right)
+
\mathcal{O}(\epsilon^2, \epsilon \theta, \theta^3)
\ef.
\ee
This proves that the expressions $\Qpm(z_i,z_j)$, both in the case
$i=j$ and $i \neq j$, are even in $\theta_j$ at leading orders, 
so we can drop out odd terms in the one-body measure 
$\mathrm{d}\theta_j \mu(\theta_j)$, when $|\theta_j| < \delta$.

This analysis holding in an interval of small $\theta$
does not mean that the remaining part of the integral
is negligible, and in general this is not the
case.  However, we can keep control on our expressions by considering
an exact subdivision of the contribution of
a diagram (\ref{eq.ID}). Indeed, in general, for a contour integration
on a path $\gamma$, and two points $a$ and $b$ on it, we can write
$\oint_{\gamma} \dx{z} f(z) = 
\int_{\gamma(a \to b)} \dx{z} f(z)
+ \int_{\gamma(b \to a)} \dx{z} f(z)$. For our angular integrations,
we can thus distinguish between the small-$\theta$ and large-$\theta$
behaviour by choosing
to decompose the (periodic) interval $[0, 2\pi]$ into
$[-\delta, \delta]$ and $[\delta, 2 \pi - \delta]$, and denote by 
$V' \subseteq V(D)$ the set of angular variables for which we
integrate in the first interval:
\begin{align}
\label{eq.IDspl}
\mathcal{I}(D) 
&=
\sum_{V' \subseteq V(D)}
\mathcal{I}(D; V')
\ef;
\\
\label{eq.64765b}
\mathcal{I}(D; V')
&=
\prod_{j \in V'} \int_{-\delta}^{\delta}
\frac{\dx{\theta_j}}{2 \pi}
\prod_{j \in V(D) \smallsetminus V'} \int_{\delta}^{2 \pi - \delta}
\frac{\dx{\theta_j}}{2 \pi}
\;
\prod_j \mu(\theta_j)
\prod_{(ij) \in E(D)}
f(\theta_i, \theta_j)
\ef.
\end{align}
Remark that, while each $\mathcal{I}(D; V')$ in (\ref{eq.64765b}) is a
function of $\delta$, the sum $\mathcal{I}(D)$ in (\ref{eq.IDspl}) is
independent from its value. This arbitrariness will be exploited in
the next paragraphs.

Both the one-body function $\mu(\theta_j)$ and the Green function
$f(\theta_i,\theta_j)$ are of order 1 if the corresponding angles are
larger than $\delta$. In particular, $\mathcal{I}(D; \emptyset)$ gives
a contribution of order 1, analytic in a
neighbourhood of $\epsilon = 0$.  Thus it should be neglected even if
it were the leading summand, in a way similar to what was shown to occur
for the 1- and 2-component cases.

The function $\mu(\theta_j)$ is given by (\ref{eq.measthet}) for
$\theta_j$ small, and thus gives ``small'' factors, of which we can
give a dimensional estimate 
\be
\int_{-\delta}^{\delta}
\frac{\dx{\theta_j}}{2 \pi}
\mu(\theta_j) \sim 
\delta (\epsilon + \delta^2)
\ef.
\ee
Factors $f(\theta_i, \theta_j)$ occur for
non-contractible arcs, with equal indices, and for bridges, with
distinct indices. In the first case, if $\theta$ is small
we get the expression in (\ref{eq.cheso}) to the power
$-\frac{1}{2}$ (plus a subleading contribution of order $1$):
\be
\label{eq.fjj}
f(\theta_j, \theta_j)
\simeq
\frac{1}{ \sqrt{\frac{2}{h}} } 
\;
\frac{1}{2 \sqrt{\epsilon + \frac{h(h+1)}{2} \theta_j^2}}
\sim (\epsilon + \delta^2)^{-\frac{1}{2}}
\ef.
\ee
In the case of bridges, we have two non-trivial cases, depending if
only one angle (say, $\theta_i$) is small, or both angles are small.
In the first case, one can neglect the contribution of the small angle
in the combination, as it does not produce any leading singularity,
and write
\be
\label{eq.67896797}
f(\theta_i, \theta_j)
\simeq 
\frac{1}{2}
\left(
\frac{ \sqrt{z_{\jind}^2-4} }{z_j-2} \pm 1
\right)
=
\frac{1}{2}
\left(
\sqrt{ \frac{z_j+2}{z_j-2} } \pm 1
\right)
(1 + \mathcal{O}(\sqrt{\epsilon}, \delta))
\ee
where the leading expression
is of order 1 by our previous assumption that 
$j \not\in V'$.

The result above
seems to suggest that, if instead both angles are small, a singular
behaviour may arise, similarly to (\ref{eq.fjj}).
Indeed, in this case, we have for the leading part in 
$f(\theta_i, \theta_j)$
\be
\frac{1}{2}
\frac{ \sqrt{z_i^2-4} - 
\sqrt{z_{\jind}^2-4} }{z_i - z_j}
\simeq
\frac{1}{2}
\frac{ \sqrt{4 (z_i-2)} - 
\sqrt{4 (z_{\jind}-2)} }{(z_i-2) - (z_j-2)}
=
\frac{ 1 }{\sqrt{z_i-2} + \sqrt{z_{\jind}-2} }
\ee
so that we obtain
\be
\label{eq.fij}
f(\theta_i, \theta_j)
\simeq
\frac{1}{ \sqrt{\frac{2}{h}} } 
\;
\frac{1}{
\sqrt{\epsilon + \frac{h(h+1)}{2} \theta_i^2}
+
\sqrt{\epsilon + \frac{h(h+1)}{2} \theta_j^2}
}
\sim (\epsilon + \delta^2)^{-\frac{1}{2}}
\ef.
\ee
Remark how the expression (\ref{eq.fjj}) is a special case of
(\ref{eq.fij}), with no need anymore of l'H\^opital limit.
At the end, we get
\be
\label{eq.IDVdim}
\mathcal{I}(D; V')
\sim
\big( \delta (\epsilon + \delta^2) \big)^{|V'|}
(\epsilon + \delta^2)^{-\frac{1}{2} \cdot \# \{ (ij) \in E(D) : i,j \in V' \} }
\ef.
\ee
The expression above is just a crude ``dimensional''
analysis. However, it allows us to understand which terms, at fixed
$n$ and in the double sum over $D$ and $V'$, are dominant in the
$\epsilon \to 0$ limit. At this aim, choose 
$\sqrt{\epsilon} \lesssim \delta \ll 1$, and call $E'(D,V')$ the set
of edges in the subgraph of $D$ induced by $V'$. Then
(\ref{eq.IDVdim}) becomes
\be
\label{eq.IDVdim2}
\mathcal{I}(D; V')
\sim
(\sqrt{\epsilon})^{3 |V'| - |E'|}
\ef.
\ee
We call \emph{order} of a pair $(D,V')$ the integer
$|E'| - 3 |V'|$. Pairs with maximum order give the leading
contribution. After some reflection, one understand which families at
fixed $|V'|$ minimize the order (besides the trivial case $|V'|=0$,
which gives order 0):
\begin{proposition}
\label{prop.111}
All and only the pairs $(D,V')$ minimizing $|E'| - 3 |V'|$
at fixed $n$ and $\tilde{n}=|V'|$
have the following defining properties:
\begin{itemize}
\item The diagram $D$ is a triangulation;
\item All vertices in $V(D) \smallsetminus V'$ have degree 1.
\end{itemize}
\end{proposition}
Remark that the converse of the second claim, that all vertices of
degree 1 are in $V(D) \smallsetminus V'$, is not true. However, this
case leads to terms which are strongly subleading in the fixed-$n$
(and arbitrary $\tilde{n}$) analysis, and trivially reabsorbed in the
contributions from smaller values of $\tilde{n}$. So, with an abuse of
definition for the class of leading diagrams, which is justified at
the light of the forthcoming equation (\ref{eq.blackleaves}), we will
restrict our attention to diagrams in which all vertices of degree 1
are in $V(D) \smallsetminus V'$.

For these diagrams it follows easily that,
for all the pairs $(D,V')$ as above,
\begin{itemize}
\item The subgraph $D'' \subseteq D$ induced by $V'$, and with loops
  dropped out, must consist of triangles and consecutive multiple edges;
\item For each pair of consecutive multiple edges in $D''$, incident on $i$ and
  $j \in V'$, a non-zero number of vertices not in $V'$ must be
  attached to $i$ in the cyclic order between the two bridges, and
  none attached to $j$, or vice versa. The relative portion of the
  original diagram must be a triagulation.
\item No vertices nor edges in $D \smallsetminus D''$ exist besides the ones
  described above.
\end{itemize}

\begin{figure}[t]
\begin{center}
\setlength{\unitlength}{25pt}
\begin{picture}(12.5,5)
\put(0,0.3){
\includegraphics[scale=1.1, bb=0 150 285 260, clip=true]{figs2.eps}}
\end{picture}
\mycaption{\label{fig.diagrsaturi}Two pairs $(D,V')$ giving the
  leading contribution in the
  set at $n$ and $|V'|$ fixed. Vertices in $V'$ and not in $V'$ are
  denoted respectively by black and white bullets, and edges in the
  auxiliary graph $D''$ are in bold.}
\end{center}
\end{figure}

\noindent
Cfr.~figure \ref{fig.diagrsaturi} for some examples.
Pairs of this form have order $2(n-3-|V'|)$, if 
$1 \leq |V'| \leq n$. Remark in particular how this formula is
in agreement with the special case $V'=V(D)$ and $D$ a triangulation,
where $|E'|=|E(D)|=3n-6$.

A simple proof of Proposition \ref{prop.111} is as follows.  Of course
we have exactly $n - \tilde{n} > 0$ vertices in $V(D) \smallsetminus
V'$. As $D$'s are connected, each of these vertices have degree at
least 1, and there exists at least one of these vertices (say, $i$),
adjacent to at least one vertex of $V'$ (say, $j$). In analogy with
the drawings in figure~\ref{fig.diagrsaturi}, we call \emph{white} and
\emph{black} the vertices respectively in $V(D) \smallsetminus V'$ and
in $V'$, so that there must exist some $i$ white, adjacent to a black
vertex $j$.

If $i$ has degree larger than 1, then we will see that $D$ cannot have
maximal order, because we can build a locally modified graph $D'$
which improves the order while remaining in the proper $(n,\tilde{n})$
class. Indeed, there is no loop in $D$ adjacent to $j$ and surrounding
$i$ alone. Build $D'$ as follows: remove all edges incident on $i$
except for $(ij)$, add the loop on $j$ surrounding $i$ alone (this
increases the order), then add a subset of edges between $j$
and the other previous neighbours of $i$, up to make $D'$ connected
(this can only further increase the order). So we get that all white
vertices neighbours to at least one black vertex, have overall degree
1. As a consequence, all white vertices are leaves, and the second
part of the statement is proven.

For the first part, now consider a face of the diagram $D$. It visits
a number of vertices, in cyclic order, possibly with repetitions.
We have just seen that in this sequence we cannot have two consecutive
white elements. So, if the face has 4 sides or more, there must be at
least two non-adjacent terminations corresponding to black
vertices. Then, we can add the corresponding edge, which is not
(planar) multiple (by definition of face), and is thus allowed, and
this improves the order. From this we deduce that all faces are
triangles.
\hfill $\square$

\medskip
Here we see how the conjecture $t^n Z_n \sim (t/\epsilon)^n ( A
(\epsilon^n + \cdots) + B (\epsilon^4 + \cdots) \ln \epsilon )$,
deduced from the analysis of sections \ref{sec.perttrees} and
\ref{sec.pertF2}, receives new elements. The case $V'=\emptyset$ would
give a contribution of the form $A (\epsilon^n + \cdots)$, while the
case $|V'|=1$, optimal for $n > 4$ (and marginally optimal for $n=4$), 
would give a relative factor
$\epsilon^{-(n-3-|V'|)} = \epsilon^{-(n-4)}$, compatible with the
non-analytic contribution $B (\epsilon^4 + \cdots) \ln \epsilon$, and
in accord with the fact that this contribution is related to the
thermodynamic limit, as a non-empty set $V'$ produces an integrand
which is not blind to the $\epsilon \to 0$ singularities. At this
stage of dimensional analysis, the more subtle possible presence of
logarithmic factors still does not emerge.

\section{Resummation of Pansy Diagrams}
\label{sec.pansy}

The defining characteristics of the dominant diagrams get simplified
in the case $V'$ consists of a single ``central'' vertex. All other
$n-1$ vertices have degree $1$, and are connected to the central
vertex by a single bridge. Then, we have $2n-5$ non-contractible arcs
with both terminations on the central vertex, the maximal allowed
number under the constraint that there are no consecutive multiple
edges, and producing a triangulation.  An example is on the left of
figure \ref{fig.diagrsaturi}.  We call these diagrams ``pansy''
diagrams.\footnote{Both for the clear resemblance with the example in
figure, and the fact that a triangulation on the Riemann sphere of our
kind is ``trilobate'', and some pansy species have petals collected in
three main directions.}

Note in particular that, as the adjacency matrix of these graphs is
the same for a given order $n$, the integrals $\mathcal{I}(D,V')$ are
all equal, so we just need to count the diagrams at order $n$, and
evaluate a single integral.

\begin{figure}[t]
\begin{center}
\setlength{\unitlength}{25pt}
\begin{picture}(15,5)
\put(0,0.3){\includegraphics[scale=1.1, bb=0 0 340 105, clip=true]{figs2.eps}}
\end{picture}
\mycaption{\label{fig.pansytriang}On the left, a typical pansy diagram
  with $n=8$. In the
  middle, the drawing obtained by inverting the coordinates, and
  compactifying the radial coordinate. On the right, the modification
  of this drawing which makes clear the bijection with triangulations
  of a polygon with $n-1$ sides.}
\end{center}
\end{figure}

If one inverts the diagram (in the sense of complex-coordinate
inversion $z \to 1/\bar{z}$, for a drawing in which the central vertex
is at the origin), one understands that pansy diagrams are in
bijection with the triangulations of regular polygons, an enumeration
problem again solved by Catalan numbers (cfr.~figure
\ref{fig.pansytriang}). More
precisely, there are $C_{n-3}$ triangulations of a polygon with $n-1$
vertices (contributing to $n$-forests), and a cyclic symmetry factor
$1/(n-1)$ should be included. So at order $n$, from the counting of
the diagrams we have a factor 
\be
\label{eq.ultcata}
\frac{(2n-6)!}{(n-1)!(n-3)!}
\ef,
\ee
and the integral is
\be
\mathcal{I}(D,\{0\})
=
\int_{-\delta}^{\delta}
\frac{\dx{\theta_0}}{2 \pi}
\prod_{j=1}^{n-1}
\int_{\delta}^{2 \pi - \delta}
\frac{\dx{\theta_j}}{2 \pi}
\;
\prod_j \mu(\theta_j)
\;
f(\theta_0,\theta_0)^{2n-5}
\prod_{j=1}^{n-1}
f(\theta_0, \theta_j)
\ef.
\ee
As we know that all contributions for 
$\{0\} \subseteq V' \subseteq V(D)$ are subleading, we can freely
include them, and get
\be
\label{eq.blackleaves}
\mathcal{I}(D,\{0\})
\simeq
\!\!\!\!\!
\sum_{\{0\} \subseteq V' \subseteq V(D)}
\!\!\!\!\!
\mathcal{I}(D,V')
=
\int_{-\delta}^{\delta}
\frac{\dx{\theta_0}}{2 \pi}
\prod_{j=1}^{n-1}
\int_{0}^{2 \pi}
\frac{\dx{\theta_j}}{2 \pi}
\;
\prod_j \mu(\theta_j)
\;
f(\theta_0,\theta_0)^{2n-5}
\prod_{j=1}^{n-1}
f(\theta_0, \theta_j)
\ef.
\ee
We recognize that $n-1$ integrations are all equal, and factorized, so
we can write
\be
\mathcal{I}(D,\{0\})
=
\int_{-\delta}^{\delta}
\frac{\dx{\theta}}{2 \pi}
\mu(\theta)
\,
f(\theta,\theta)^{2n-5}
\left(
\int_{0}^{2 \pi}
\frac{\dx{\theta'}}{2 \pi}
\mu(\theta')
f(\theta, \theta')
\right)^{n-1}
\ef.
\ee
The integral in parenthesis, for $h=1$, is related to the one described in
equation (\ref{eq.Phi_ser2}), with 
\be
\label{eq.defxi}
x(\epsilon, \theta)=z-2\sim\frac{2}{h} 
\left( \epsilon + \frac{h(h+1)}{2} \theta^2 \right)
\ef.
\ee
In particular, for $h=1$, at leading order it can be replaced by the
numerical
constant $K_1=3-\frac{16 \sqrt{2}}{3 \pi}$, corresponding to the limit
in which both $x$ and $\epsilon$ vanish. More generally, we expect
that this limit is finite for any value of $h$, and call it $K_h$
\be
\label{eq.defKh}
K_h = 
\oint \frac{\dx{x}}{2 \pi i}
g_c x^{h+1} (1-(h+1) g_c x^h))
\frac{ q\big( x(1-g_c x^h) \big) - 1}{x(1-g_c x^h) - 2}
\ef,
\ee
which is the limit for $w \to 2$ and $g \to g_c$ of the quantity
(\ref{eq.PhiL}) in the case $\ell=1$.

The combination $x$ above is useful, as it also appears in
$f(\theta,\theta)$
\be
f(\theta,\theta)
=
\frac{1}{2 \sqrt{x}} (1+\mathcal{O}(\epsilon,\delta))
\ef,
\ee
so it is convenient to express also $\mu(\theta)$ in these terms
\be
\mu(\theta)
=
\frac{h}{2} \frac{3h+4}{h+1} x
-
\frac{2h+3}{h+1} \epsilon
+
\mathcal{O}(\epsilon^2,\epsilon \delta, \delta^3))
\ef.
\ee
More precisely, it has the structure
\be
\mu(\theta)
=
\left(1 - \alpha \right) 
\epsilon
+ 
\frac{h \alpha}{2}
x
+
\mathcal{O}(\epsilon^2,\epsilon \delta, \delta^3))
\ef,
\ee
with
\be
\label{eq.defalpha}
\alpha = 
3+\frac{1}{h+1}
\ef,
\ee
and our relevant expression is
\be
\mathcal{I}(D,\{0\})
=
2^{-2n+5}
K_h^{n-1}
\int_{-\delta}^{\delta}
\frac{\dx{\theta}}{2 \pi}
\left(
\frac{h \alpha}{2}
x
+
\left(1 - \alpha \right) 
\epsilon
\right)
x^{-n+\frac{5}{2}}
\ef,
\ee
up to subleading terms. So we need to consider integrals of the form
\be
W_{m}(\epsilon,\delta)
=
\int_{-\delta}^{\delta}
\frac{\dx{\theta}}{2 \pi}
x(\epsilon, \theta)^{-m-\frac{1}{2}}
\ef,
\ee
for $m \geq -1$. Some scale factors overall are easily
extracted. Defining $\tilde{\delta}=\sqrt{\frac{h(h+1)}{2}} \delta$ we
have
\be
W_{m}(\epsilon,\delta)
=
\left( \frac{h}{2} \right)^m
\frac{1}{\sqrt{h+1}}
\int_{-\tilde{\delta}}^{\tilde{\delta}}
\frac{\dx{\tau}}{2 \pi}
(\epsilon + \tau^2)^{-m-\frac{1}{2}}
\ef.
\ee
The expression for $Z_n$, collecting also the combinatorial factor
(\ref{eq.ultcata}), thus reads
\be
\label{eq.goodpoint}
\begin{split}
t^n Z_n
&\simeq
t^n
\frac{(2n-6)!}{(n-1)!(n-3)!}
\mathcal{I}(D,\{0\})
\\
&=
t^n
\frac{(2n-6)!}{(n-1)!(n-3)!}
\frac{1}{2 \pi}
2^{-2n+5}
K_h^{n-1}
\left(
\frac{h \alpha}{2}
W_{n-4}
+
\left(1 - \alpha \right) 
\epsilon
\, W_{n-3}
\right)
\\
&=
\frac{t}{\pi}
\frac{(2n-7)!!}{(n-1)!}
\left( \frac{t K_h}{2} \right)^{n-1}
\left(
\frac{h \alpha}{2}
W_{n-4}
+
\left(1 - \alpha \right) 
\epsilon
\, W_{n-3}
\right)
\ef.
\end{split}
\ee
For $m=-1$ and $m=0$ (corresponding respectively to $n=3$ and $n=4$)
the integral $W_m$ is not convergent in
the limit $\tilde{\delta} \to \infty$, and contains a combination of
logarithms that we shall discuss:
\begin{align}
\int_{-\delta}^{\delta}
\dx{\tau}
(\epsilon + \tau^2)^{\frac{1}{2}}
&=
\delta \sqrt{\epsilon + \delta^2}
+ \epsilon \ln (\delta + \sqrt{\epsilon + \delta^2})
- \frac{1}{2} \epsilon \ln \epsilon
\ef;
\\
\int_{-\delta}^{\delta}
\dx{\tau}
(\epsilon + \tau^2)^{-\frac{1}{2}}
&=
2 \ln (\delta + \sqrt{\epsilon + \delta^2})
- \ln \epsilon
\ef.
\end{align}
For $m>0$ the integral is convergent.
The general formula in the limit $\delta \to \infty$ is
\be
\int_{-\infty}^{\infty}
\dx{\tau}
(\epsilon + \tau^2)^{-m-\frac{1}{2}}
=
\epsilon^{-m} \frac{2^m (m-1)!}{(2m-1)!!}
\ef.
\ee
Our dimensional analysis in powers of $\epsilon$ and $\delta$
is valid for a wide range of values
for $\delta$ (it suffices that $\delta \ll 1$, and 
$\delta \gtrsim \sqrt{\epsilon}$), so every term in the result which
depends on $\delta$ in a way which is not compatible with this arbitrariness
must be intended as coming from the analytic part of the integral,
being this a leading term, as in the cases $n=3$ and $n=4$, or a
subleading one, as in the case $n \geq 5$ with finite $\tilde{\delta}$
(as it is legitimate to take $\delta/\sqrt{\epsilon} \gg 1$). Thus, at
the aim of understanding the non-analytic contribution, these terms
can be dropped out, and we have
\be
\label{eq.Wmm}
W_m^{\rm (n.a.)}(\epsilon) =
\frac{1}{\sqrt{h+1}}
\left(
\frac{\epsilon}{h} \right)^{-m}
\times
\left\{
\begin{array}{ll}
- \frac{(-2m-1)!!}{(-m)!} \ln \epsilon
& m \leq 0 \\
\rule{0pt}{14pt}%
\frac{(m-1)!}{(2m-1)!!}
& m \geq 1
\end{array}
\right.
\ee
Substituting (\ref{eq.Wmm}) in (\ref{eq.goodpoint}), again in
agreement with our
conjecture, we obtain the expressions for $n=3$ and $n=4$, that we
report here together with the ones (for $h$ odd), already derived in equations
(\ref{eq.epsF1}) and (\ref{z2})
\begin{align}
t Z_1(g_c e^{-\epsilon})
&=
\left( \frac{t}{\epsilon} \right)
\left[
( \textrm{$\mathcal{O}(\epsilon)$ analytic, $\mathcal{O}(\epsilon^5)$} )
+
\frac{2 \sqrt{h+1}}{3 \pi h^4}
\; \epsilon^4
\ln \epsilon
\right]
\ef;
\\
t^2 Z_2(g_c e^{-\epsilon})
&=
\left( \frac{t}{\epsilon} \right)^2
\left[
( \textrm{$\mathcal{O}(\epsilon^2)$ analytic, $\mathcal{O}(\epsilon^5)$} )
-
\frac{\sqrt{h+1} \, K_h}{\pi h^3}
\; \epsilon^4
\ln \epsilon
\right]
\ef;
\\
t^3 Z_3(g_c e^{-\epsilon})
&=
\left( \frac{t}{\epsilon} \right)^3
\left[
( \textrm{$\mathcal{O}(\epsilon^3)$ analytic, $\mathcal{O}(\epsilon^5)$} )
+
\frac{(\alpha-2) \, K_h^2}{16 \pi \sqrt{h+1}}
\; \epsilon^4
\ln \epsilon
\right]
\ef;
\\
t^4 Z_4(g_c e^{-\epsilon})
&=
\left( \frac{t}{\epsilon} \right)^4
\left[
( \textrm{$\mathcal{O}(\epsilon^3)$ analytic, $\mathcal{O}(\epsilon^5)$} )
-
\frac{\alpha\,h \, K_h^3}{96 \pi \sqrt{h+1}}
\; \epsilon^4
\ln \epsilon
\right]
\ef.
\end{align}
For $n \geq 5$, as the overall power of $\epsilon$ is negative,
the pansy diagrams give a non-analytic quantity (in a neighbourhood
of $\epsilon=0$) already without a logarithmic factor, which indeed
does not occur.

Substituting (\ref{eq.Wmm}) in (\ref{eq.goodpoint}), 
we get a series for the terms with $n \geq 5$ 
\be
\label{eq.fullsu}
\sum_{n \geq 5}
t^n Z_n(g_c e^{-\epsilon})
\simeq
\frac{(t/\epsilon) \, \epsilon^4}{\pi \, h^2\,\sqrt{h+1}}
\sum_{n \geq 5}
\left(
\frac{t}{\epsilon}
\frac{h K_h}{2}
\right)^{n-1}
\left(
\frac{(n-4)!}{(n-1)!}
+
\frac{\alpha}{2}
\frac{(n-5)!}{(n-1)!}
\right)
\ef.
\ee
The two series are trivial, and have radius of convergence 1. The
behaviour near to this point is deducible from the exact expressions
\begin{align}
\sum_{n \geq 5}
\left(
1-x
\right)^{n-1}
\frac{(n-4)!}{(n-1)!}
&=
-\frac{1}{2} x^2 \ln x
+\frac{1}{12} (1-x)(1-5x-2x^2)
\ef;
\\
\sum_{n \geq 5}
\left(
1-x
\right)^{n-1}
\frac{(n-5)!}{(n-1)!}
&=
\frac{1}{6} x^3 \ln x
+\frac{1}{36} (1-x)(2-7x+11 x^2)
\ef.
\end{align}
Remark in particular the leading non-analytic behaviour in the full
sum (\ref{eq.fullsu}), for $\epsilon \searrow \frac{t h K_h}{2}$
\be
\sum_{n \geq 5}
t^n Z_n(g_c e^{-\epsilon})
\simeq
-
\frac{\epsilon^4}{\pi K_h h^3 \sqrt{h+1}}
\left(
1 -
\frac{t}{\epsilon}
\frac{h K_h}{2}
\right)^2
\ln 
\left(
1 -
\frac{t}{\epsilon}
\frac{h K_h}{2}
\right)
\ef,
\ee
regardless of the precise expression (\ref{eq.defalpha}) for $\alpha(h)$.
We can interpret the result of this calculation as describing the
curve of critical values $g_c(t;h)$ in the theory of Spanning Forests,
in a neighbourhood of $t=0$, when the limit $\eval{ |V(G)| } \to
\infty$ has been taken before $K(F) \to \infty$.
We already know from elementary means the
formula for $g_c(0;h)$, equation (\ref{eq.gcrit}). The result above
states that, in the limit described above,
\be
\left.
\frac{\dx{}}{\dx{t}} \ln g_c(t;h) \right|_{t=0}
=
- \frac{h K_h}{2}
\ef,
\ee
with $K_h$ a numerical constant, obtained from a single (non-singular)
one-dimensional contour integral, in equation~(\ref{eq.defKh}), and
known for $h=1$.

The whole analysis of the pansy diagrams, besides being a remarkable
exact result, shows a ``topological'' fact of this peculiar large-volume
limit, namely that, for $t$,
$\epsilon$ small and 
\be
\label{eq.defestar}
\epsilon 
\searrow
\epsilon_*(t)
:=
\frac{t h K_h}{2}
\ef,
\ee
the partition sum is dominated by forests whose adjacency diagram is
compatible with the presence of a single gigantic tree, and many small
trees. Although this statement is not quantitative, it has a clear
topological reformulation: at order $K(F)=n$ fixed and in the limit
$\eval{|V(G)|} \to \infty$, almost surely there is a single tree $T$
which is neighbour of any other tree $T'$, and, for each $T'$, there
are edges in $G \smallsetminus F$ with both endpoints on $T$, such
that the edge, together with the unique path connecting the endpoints
on $T$, makes a cycle which encircles $T'$.  At the same time, no
pairs $T'$, $T''$ of trees distinct from $T$ are adjacent.

\section{Conclusions and perspectives}
\label{sec.moregigs}

The statistical mechanics of spanning forests on various graphs has
two main kinds of criticality, in the ``probabilistic regime'' of $t$
real non-negative, besides the trivial ``high temperature'' fixed
point $t \to +\infty$.  The point $t=0$ corresponds to a massless
theory of a scalar fermion, at the light of Kirchhoff Matrix-Tree
theorem. Furthermore, at some critical value $t^{\star}$, a
\emph{percolation} transition may occur, i.e.~for values $t <
t^{\star}$ there exist trees which occupy a fraction of order 1 of
the volume (\emph{gigantic} compents), while for $t > t^{\star}$ all
trees have a characteristic size, depending on $t$ alone and not
scaling with the volume. This is the specialization to forests of a
feature holding more generally for the probabilistic sector of the
Random Cluster (Potts) model.  It is shown numerically in three, four
and five dimensions \cite{alanragazzi}, and analytically in the
``infinite-dimensional'' limit of fully connected graphs
\cite{noibedini} that $t^{\star}$ is finite and non-zero in these
cases, while it is expected that the arising asymptotic freedom for
the model of spanning forests in two Euclidean dimensions is due to
the fact that the ``Kirchhoff'' criticality and the percolation
criticality do coincide exactly for $d=2$.

Indeed, it was also at the aim of understanding rigorously this set of
conjectures, that we performed the study of the model on Random Planar
Graphs, with the aim of combining the results with KPZ tools.

We have however to face a problem in the interpretation of the results
of the previous section. As we said, we have performed a double limit of
$\eval{ |V(G)| } \to \infty$ and $K(F) \to \infty$, where the first
one has been performed before the second one. On the contrary, at
least on Euclidean lattices, for any
finite $t$, we expect a macroscopic number of components, 
i.e.~$\eval{ \frac{K(F)}{|V(G)|} } = \mathcal{O}(1)$, and, in order to
have a better understanding on the behaviour of the system, we would like to
control the double limit $\eval{ |V(G)| }, K(F) \to \infty$ with arbitrary scaling.

We have seen how, in the Feynman expansion, the accessory parameter
$V'$ has a deeper meaning than it was legitimate to expect: when, for
a given diagram, $V'$ is chosen in order to maximize the
contribution, we have that vertices respectively in $V'$ and not,
correspond to gigantic and small trees. So we expect that a control on
the microcanonical ensemble for $\tilde{n}:= |V'|$ would help at the
aim of understanding the various limits. This is in a sense
generalizing the approach of the previous section, where we stated
that $\tilde{n}=0$ must give an analytic contribution, coming from
graphs of small size, and we analyzed exactly the leading contribution
to $\tilde{n}=1$ for $g \to g_c$.

Proposition~\ref{prop.111} already identifies the class of  leading diagrams at both $n$ and $\tilde{n}$ fixed.
We plan in the next future to attack the problem of  re-summation of diagrams in $n$  at  further values of $\tilde{n}$.

\appendix
\section{Generating functions for $k$-trees and Hypergeometric
  functions}
\label{app.HG}

We recall the definitions (\ref{eq.adef}, \ref{eq.a'def})
\begin{align*}
A_{h,n}
&=\frac{((h+1)n)!}{n! (hn+1)!}
\ef;
&
A_h(\omega)
&=\sum_{n\geq 0} \omega^n A_{h,n}
\ef;
\\
A'_{h,n}
&=\frac{((h+1)n)!}{n! (hn+2)!}
\ef;
&
A'_h(\omega)
&=\sum_{n\geq 1} \omega^n A'_{h,n}
\ef.
\end{align*}
From the ratio of two consecutive summands
\begin{align}
\frac{A_{h,n+1}}{A_{h,n}}
&=
\frac{(h+1)^{h+1}}{h^{h}}
\,
\frac{\left( n+\frac{h}{h+1} \right) \cdots
\left( n+\frac{1}{h+1} \right)}
{\left( n+\frac{h+1}{h} \right) \cdots
\left( n+\frac{2}{h} \right)}
\end{align}
and
\begin{align}
\frac{A'_{h,n+1}}{A'_{h,n}}
&=
\frac{(h+1)^{h+1}}{h^{h}}
\,
\frac{\left( n+\frac{h}{h+1} \right) \cdots
\left( n+\frac{1}{h+1} \right)}
{\left( n+\frac{h+2}{h} \right) \cdots
\left( n+\frac{3}{h} \right)}
\end{align}
and the definition of generalized Hypergeometric 
function~\cite{hyperg}
\begin{gather}
{}_p F_q ({\bf a}; {\bf b}; \omega)
:= \sum_{n=0}^{\infty} \alpha_n \omega^n
\ef;
\\
\frac{\alpha_{n+1}}{\alpha_n}=\frac{(n+a_1) \cdots (n+a_p)}
{(n+b_1) \cdots (n+b_q) (n+1)}
\ef;
\\
\alpha_0=1
\ef;
\end{gather}
we have
\begin{subequations}
\label{eq.acomehg}
\begin{align}
A_h(\omega) &= {}_{h+1}F_{h}({\bf a}_h; {\bf b}_h; c_h \omega)
\ef;
\\
{\bf a}_h &= \left(
\smfrac{1}{h+1}, \ldots, \smfrac{h}{h+1} \right)
\ef;
\\
{\bf b}_h &= \left(
\smfrac{2}{h}, \ldots, \smfrac{h-1}{h}, 
\smfrac{h+1}{h} \right)
\ef;
\\
c_h &= \frac{(h+1)^{h+1}}{h^{h}}
\ef;
\end{align}
\end{subequations}
and
\begin{subequations}
\begin{align}
A'_h(\omega) &=\smfrac{1}{2} \left( 
{}_{h+1}F_{h}({\bf a}_h; {\bf b'}_h; c_h \omega) -1
\right)
\ef;
\\
{\bf b'}_h &= \left(
\smfrac{3}{h}, \ldots, \smfrac{h-1}{h}, 
\smfrac{h+1}{h}, \smfrac{h+2}{h} \right)
\ef.
\end{align}
\end{subequations}
The parameters of functions $A_h$ and $A'_h$ have
all but one entry in common, the last one differing by 1, so they are
\emph{contiguous} as hypergeometric functions (cfr.~\cite{hyperg},
par.~2.2.1), and this accounts for the simple relation (\ref{eq.aa'}).

The hypergeometric function corresponding to $A_h(\omega)$ has been
already studied by M.L.\ Glass\-er\footnote{M.L.~Glasser, pers.~comm.~to
  the Wolfram Mathworld community, Sept.~26, 2003\\
{\tt http://mathworld.wolfram.com/HypergeometricFunction.html}
eq.~25.}. In particular, the relation (25) in his communication
coincides with our parametric solution (\ref{eq.solparam}), after that
$A_h(\omega)$ has been identified with the hypergeometric function in 
(\ref{eq.acomehg}).

The hypergeometric functions ${}_pF_q$ 
with $p=q+1$ and non-integer parameters do not have poles and
essential singularities. They have a branch-cut discontinuity between
$z^*=1$ and the point at infinity. Accounting for the rescaling constant
$c_h$, we have
$\omega^* = h^{h}/(h+1)^{h+1}= 2^h g_c(h)$ 
(cfr.~definition (\ref{eq.gcrit})), and in particular
$\omega^* = 1/4$ for $h=1$, as clear from the explicit expressions
(\ref{eq.a1}) and~(\ref{eq.a'1}).

\section{Comparison with random spanning trees}

The results above for the ensemble of random planar lattice can be
compared with small effort with the case of random lattices,
regardless to the genus. In this case, the only difference is that the
Catalan numbers, deriving from the combinatorics of planar matchings,
must be replaced with the number of arbitrary matchings of $2n$
points, which are $(2n-1)!!$. Thus, provided that $hV$ is even, we have
\be
Z_1(g)= \sum_V g^V 
A'_{h,V} 
(hV+1)!!
\ef.
\ee
Again we treat separately the two cases of $h$ odd or even.
In the case of $h$ odd we have
\be
\label{eq.47835rnd}
Z_1(g)= \sum_V g^{2V} \frac{(2V(h+1))!}{(2V)! (hV+1)!\, 2^{hV+1}}
\ef,
\ee
from which we have the asymptotics
\be
Z_1 (g) \sim \sum_V
(hV)! \left( g \frac{(h+1)^{h+1}}{(h^2/2)^{h/2}} 
\right)^{2V}\!\! V^{-2}
\ef,
\ee
while in the case of $h$ even
we have the formula
\be
\label{eq.47835bisrnd}
Z_1(g)= \sum_V g^{V} \frac{(V(h+1))!}
{V! (\smfrac{1}{2} hV+1)!\, 2^{hV/2+1}}
\ef,
\ee
from which we have the asymptotics
\be
Z_1 (g) \sim \sum_V
(\smfrac{1}{2} hV)! \left( g \frac{(h+1)^{h+1}}{(h^2/2)^{h/2}} 
\right)^{V}\!\!\! V^{-2}
\ef.
\ee
Remark the expected ``entropic catastrophe'', due to the
super-exponential number of random lattices. As a consequence, the
related hypergeometric function is a ${}_{q} F_p$,
with $q-p>1$, contrarily to what happens in
the planar case, in which one deals with ${}_{p+1} F_p$ functions,
which have a finite radius of convergence.


\end{document}